\documentclass[journal]{IEEEtran}

\usepackage[english]{babel}
\usepackage{graphicx}
\usepackage{xcolor}
\usepackage{amsmath}
\usepackage{amssymb}
\usepackage{bm}
\usepackage{stmaryrd}
\usepackage[normalem]{ulem}
\usepackage{tabularx}
\usepackage{booktabs}
\usepackage{hyperref}
\usepackage{cite}
\usepackage{subcaption}
\usepackage{xspace}
\usepackage[T1]{fontenc}

\usepackage{algorithm}
\usepackage{algpseudocode}
\algrenewcommand\algorithmicrequire{\textbf{Inputs:}}
\algrenewcommand\algorithmicensure{\textbf{Outputs:}}
\algnewcommand{\LineComment}[1]{\Statex \hfill {\scriptsize$\triangleright$ #1}}
 
\newcommand{\blue}[1]{\textcolor{black}{#1}}

\newcommand{\cb}{\color{black}} 
\newcommand{\cblue}{\color{black}}

\newcommand{\para}[1]{\smallskip\noindent{\bf #1}}
\newcommand{\spara}[1]{\smallskip\noindent{\bf #1}}
\newcommand{\softpara}[1]{\smallskip\noindent\underline{#1}}

\newcommand{\dstt}{\mbox{\tt DST}\xspace}
\newcommand{\dst}{\mbox{\tt DST}\xspace}
\newcommand{\scope}{\mbox{SCOPE}\xspace}
\newcommand{\eh}{\mbox{\tt E2H}\xspace}
\newcommand{\me}{\mbox{\tt EM}\xspace}
\newcommand{\shp}{\mbox{\tt SP}\xspace}

\newcommand{\hpredi}{\mathbf{h}^{(i)}_{\mathrm{pred}}}

\let\emph\textit

\newcommand{\eat}[1]{}

\newcounter{packednmbr}
\newenvironment{packedenumerate}{
\begin{list}{\thepackednmbr.}{\usecounter{packednmbr}
\setlength{\itemsep}{0pt}
\setlength{\labelwidth}{8pt}
\setlength{\leftmargin}{12pt}
\setlength{\labelsep}{4pt}
\setlength{\listparindent}{\parindent}
\setlength{\parsep}{0pt}
\setlength{\topsep}{1pt}}}{\end{list}}

\newenvironment{packeditemize}{
\begin{list}{$\bullet$}{
\setlength{\itemsep}{1.5pt}
\setlength{\labelwidth}{8pt}
\setlength{\leftmargin}{10pt}
\setlength{\labelsep}{3pt}
\setlength{\listparindent}{\parindent}
\setlength{\parsep}{1.5pt}
\setlength{\parskip}{1.5pt}
\setlength{\topsep}{1.5pt}}}{\end{list}}

\usepackage{titlesec}

\titleformat{\section}[block]
  {\normalfont\normalsize\bfseries\centering}
  {\thesection.}{1em}{\MakeUppercase}

\renewcommand{\thesubsection}{\Alph{subsection}}
\titleformat{\subsection}
  {\it\normalsize\bfseries}
  {\thesubsection.}{1em}{}

\begin{document}

\title{SCOPE: A Syndrome-Driven Control Plane for\\
QEC-Enabled Quantum Networks}

\author{Xiaojie~Fan,~Zian~Wang,~Ashutosh~Tiwari,
  ~Himanshu~Gupta
  
  \IEEEauthorblockA{\textit{Department of Computer Science, Stony Brook University, Stony Brook, NY 11790, USA}}
  \thanks{Corresponding author: Xiaojie Fan (email: xiffan@cs.stonybrook.edu)}
}

\maketitle

\begin{abstract}
  As quantum networks evolve from experimental testbeds to fault-tolerant systems, the primary performance metric shifts from physical link fidelity to end-to-end logical error rate.
However, current control planes remain ill-equipped for this transition: routing decisions are typically decoupled from Quantum Error Correction (QEC) strategies, relying on topology or scalar fidelity metrics that fail to predict how specific physical noise structures interact with logical codes.
Optimizing this coupled route-and-code performance requires precise, real-time visibility into network error biases, yet traditional active tomography is operationally prohibitive due to throughput collapse and service interruption.

We present \scope (Syndrome-based COntrol PlanE), a network-layer architecture that enables joint routing and coding optimization using purely passive telemetry.
Instead of injecting probes, \scope harvests \emph{error syndromes}—the parity check outcomes naturally generated by QEC decoders during user service.
By aggregating these signals, SCOPE’s inference engine reconstructs the network’s time-varying error map, capturing complex, context-dependent noise correlations. 

This visibility drives a decision engine that proactively pushes optimal route-and-code configurations to source nodes.
NetSquid and IBM-calibrated simulations show that \scope reduces estimation error by more than 60\% relative to a standard EM baseline. In large-scale networks, this precision reduces logical error rates by 30--35\% (up to 65\%) against topology-aware baselines.

\end{abstract}

\begin{IEEEkeywords}
Quantum networks, quantum error correction, syndrome
tomography, software-defined networking, joint routing
and coding
\end{IEEEkeywords}

\section{Introduction}
\label{sec:intro}

Quantum networks are poised to evolve from small-scale physics experiments into planetary-scale distributed systems.
While early prototypes rely on probabilistic schemes such as entanglement purification (Generation-1), scalable operation demands determinism.
Consequently, the field is shifting toward \emph{Quantum Error Correction (QEC)}-enabled networks (Generation-2/3), where information is protected logically rather than just physically~\cite{muralidharan2016optimal}.
In this regime, the primary objective of the network stack changes fundamentally~\cite{nickerson2013topological, wehner2018quantum}: it is no longer to maximize raw entanglement fidelity but to minimize the \emph{End-to-End Logical Error Rate (LER)}.

This shift exposes a critical gap in the control plane.
Existing quantum routing protocols typically treat the network as a static graph, optimizing for topological (shortest-path~\cite{van2013path}) or scalar physical metrics (e.g., fidelity~\cite{shi2020redundant}).
We argue that this approach is insufficient for QEC networks because it decouples routing from coding.
In reality, LER is a non-linear function of both the path's specific noise structure (e.g., $Z$ or $X$ error bias) and the QEC code's ability to suppress that structure.
A ``high-fidelity'' path with $Z$-biased noise might be disastrous for a code with a low $Z$-threshold, whereas a ``lower-fidelity'' path with uniform noise might yield a higher logical throughput.
Thus, an effective control plane must perform \textbf{Joint Route-Code Optimization}, identifying the (route, code) combination that minimizes the LER.

Implementing such a control plane faces two fundamental challenges: (i) \emph{Visibility}: acquiring a detailed, up-to-date map of Pauli error rates ($p_X, p_Y, p_Z$) for every link, and (ii) 
\emph{Modeling:} accurately mapping isolated link errors to end-to-end path performance in the presence of correlations.
Traditionally, acquiring visibility requires \emph{Active Tomography}---halting application traffic to inject dedicated probe states. In production environments, this is operationally prohibitive; it necessitates halting user traffic, competes for scarce resources, and fails to capture the load-dependent noise (e.g., crosstalk) present during active service.
Furthermore, active probes characterize links in isolation, forcing the control plane to estimate end-to-end performance based on static, independence assumptions. In reality, quantum errors are often correlated across links~\cite{pfister2018capacity} (e.g., due to shared control signal); thus, even a perfect set of link tomographs yields an incorrect prediction of the LER.

In this paper, we present \scope (Syndrome-based COntrol PlanE), a system that leverages {\bf passive syndrome tomography} to drive optimal networking.
Our key insight is that QEC decoders naturally generate continuous telemetry in the form of \emph{syndromes}---parity measurements intended for local correction that actually contain the statistical ``fingerprint'' of the path's noise environment.
\scope harvests these signals from live user traffic to continuously reconstruct the physical error map without injecting a single probe.
This closes the network control loop: the system ingests syndromes from end-points, continually learns a network-wide error profile, and dynamically pushes optimized Routing and Coding configurations.
To handle the complexity of quantum noise (often correlated and non-Markovian), \scope employs a hybrid learning architecture: a fast analytical engine handles independent errors, while a deep-learning correlation-aware engine captures complex, path-dependent correlations.
We make the following contributions:
\begin{packedenumerate}
    \item \emph{The \scope Framework:} We propose a novel system architecture that leverages \emph{passive syndrome tomography}—using error correction data from live user traffic—to estimate network error profiles in real-time. 
    
    \item \emph{Differentiable Syndrome Tomography (\dst):} For independent error models, we develop the \dst engine. By formulating syndrome generation as a differentiable forward model, \dst solves the inverse problem via gradient descent to recover link-specific Pauli error rates $(p_x, p_y, p_z)$. 
    
    \item \emph{Correlation-Aware Learning Engine:} To address spatially or temporally \emph{dependent} error models, we introduce a deep learning engine built on Graph Neural Networks (GNNs) and Transformers. This engine captures complex non-linear correlations in noise patterns, enabling high-fidelity estimation in general error models.
    
    \item \emph{Generalization to Physical Constraints:} We extend the core estimation techniques to handle practical hardware realities, including: \emph{decoherence errors}, teleportation-based transmissions, and support for heterogeneous QEC codes.
    
    \item \emph{Joint Route-Code Selection:} Leveraging the learned error profiles, we formulate and solve the \emph{Joint Route-Code Selection} problem. Unlike standard routing, which optimizes distance or raw fidelity, our optimizer selects the route and QEC code combination that minimizes the estimated \emph{logical error rate}, exploiting code-bias asymmetries.
    
    \item \emph{Evaluation:} NetSquid and IBM-calibrated simulations show that \scope reduces estimation error by more than 60\% relative to a standard EM baseline. In large-scale networks, this precision reduces logical error rates by 30--35\% (up to 65\%) against topology-aware baselines.  
\end{packedenumerate}

\section{Background}
\label{sec:background}

\para{Quantum Network (QN) Abstraction.}
A quantum network is a distributed system of quantum-capable nodes connected by optical channels.
Formally, we model a QN as a graph $G = (V, E)$, where $V$ represents nodes equipped with quantum memories and local logic, and $E$ represents lossy quantum links (e.g., optical fiber) and parallel classical control channels.
Unlike classical networks where links are static, quantum links are dynamic resources: edges in $E$ typically represent entanglement resources (Bell pairs) generated stochastically or optical paths for qubit transmission.

\para{Quantum Error Correction (QEC).}
Quantum data is fragile. To transmit information reliably over this noisy graph, we rely on Quantum Error Correction (QEC).

\softpara{Physical Errors and Pauli Noise.}
The fundamental unit of noise is modeled using the Pauli group: $I$ (identity), $X$ (bit-flip), $Z$ (phase-flip), and $Y$ (combined bit-phase flip).
While physical noise in optical fibers and memories is continuous (e.g., amplitude damping or dephasing), standard randomized compiling techniques (like Pauli Twirling) allow us to approximate these channels as a discrete \emph{Pauli Channel}.
A link is thus characterized by a probability vector $\vec{p} = (p_X, p_Y, p_Z)$~\cite{nielsen2000quantum}.
In real hardware, this vector is rarely uniform; device physics often induces strong \emph{biases}~\cite{tuckett2018ultrahigh} (e.g., $p_Z \gg p_X$ in dephasing-limited memories) or spatial correlations that significantly affect how errors accumulate.

\softpara{Logical Encoding.}
QEC protects information by encoding $k$ logical qubits into a larger block of $n$ physical qubits, denoted as an $[[n,k,d]]$ code.
The parameter $d$ (distance) quantifies fault tolerance: the code typically corrects any error pattern affecting up to $\lfloor (d-1)/2 \rfloor$ physical qubits.
Crucially, this correction capability can be biased, meaning the logical qubit may suppress specific error types (e.g., phase flips) more effectively than others depending on the code structure.
Common code families include the \emph{Surface Code} (high threshold, planar geometry) or \emph{Repetition Codes} (specialized for biased noise).
The choice of code defines a ``Logical Qubit'' and transforms the physical error rates of the path into a much lower LER.

\softpara{Syndromes: The Parity Signal.}
QEC is designed to \emph{detect} errors without measuring (and thus collapsing) the data.
This is achieved via \emph{stabilizers}---parity checks performed on subsets of physical qubits.
Intuitively, a stabilizer asks, ``Is the parity of this group even?'' without asking ``What are the values?''.
In a noiseless state, all checks pass (outcome 0).
When an error occurs (e.g., a bit-flip on a data qubit), it anticommutes with adjacent stabilizers, flipping their outcome to 1.
The vector of all stabilizer outcomes constitutes the syndrome ($\vec{s}$).
\emph{Key Property:} The syndrome is a ``many-to-one'' map.
A single physical error produces a deterministic syndrome, but a single syndrome could have been caused by many different error patterns.

\softpara{Decoding and Logical Failure.}
The \emph{decoder} addresses an inverse problem: from a measured syndrome $\vec{s}$, it selects an estimated physical error $E_{\mathrm{est}}$ (often the minimum-weight explanation) and applies the corresponding correction.
A \emph{logical error} occurs when the true error $E$ is not uniquely identifiable from $\vec{s}$ and differs from $E_{\mathrm{est}}$ by a nontrivial logical operator.
In that case, the correction returns the state to the codespace, but the net operation $E\cdot E_{\mathrm{est}}$ enacts a logical transformation (e.g., $|0\rangle \mapsto |1\rangle$), placing the state in the wrong logical sector.
Thus, minimizing logical failure requires aligning the code's protection capabilities with the specific \emph{structure} of the noise (e.g., bias), ensuring that the code can correct the most frequent error patterns.

\para{Communication Regimes (2G and 3G).}
Following Muralidharan et al.~\cite{muralidharan2016optimal}, we focus on scalable, QEC-enabled networks. In \textbf{2G}, end-to-end entanglement is created 
via \emph{Swap Trees}~\cite{ghaderibaneh2022efficient} of logical links; in 
\textbf{3G}, encoded qubits undergo direct physical transmission. \scope unifies these 
distinct data planes under a single control abstraction, the \emph{Network Plan} 
($\mathcal{P}$). We define this tuple as $\mathcal{P} = (\text{SwapTree}, 
\text{Code})$ for 2G and $\mathcal{P} = (\text{Route}, \text{Code})$ for 3G, allowing 
the router to optimize pathing and encoding jointly.

\section{Motivation: Why Cross-Layer Control?}
\label{sec:motivation}

Existing proposals for QN control typically adopt a decoupled model: the network layer selects a path based on scalar physical metrics (e.g., entanglement fidelity), while the application layer independently selects a Quantum Error Correction (QEC) code.
In this section, we argue that this decoupled approach is fundamentally flawed.
We show that logical reliability depends non-linearly on the interplay between specific noise structures and code properties, necessitating a joint approach.
Furthermore, we argue that the visibility required to drive this joint optimization cannot be achieved through traditional active tomography, thereby motivating the use of passive syndrome feedback.

\begin{figure}
    \centering
    \includegraphics[width=0.5\textwidth]{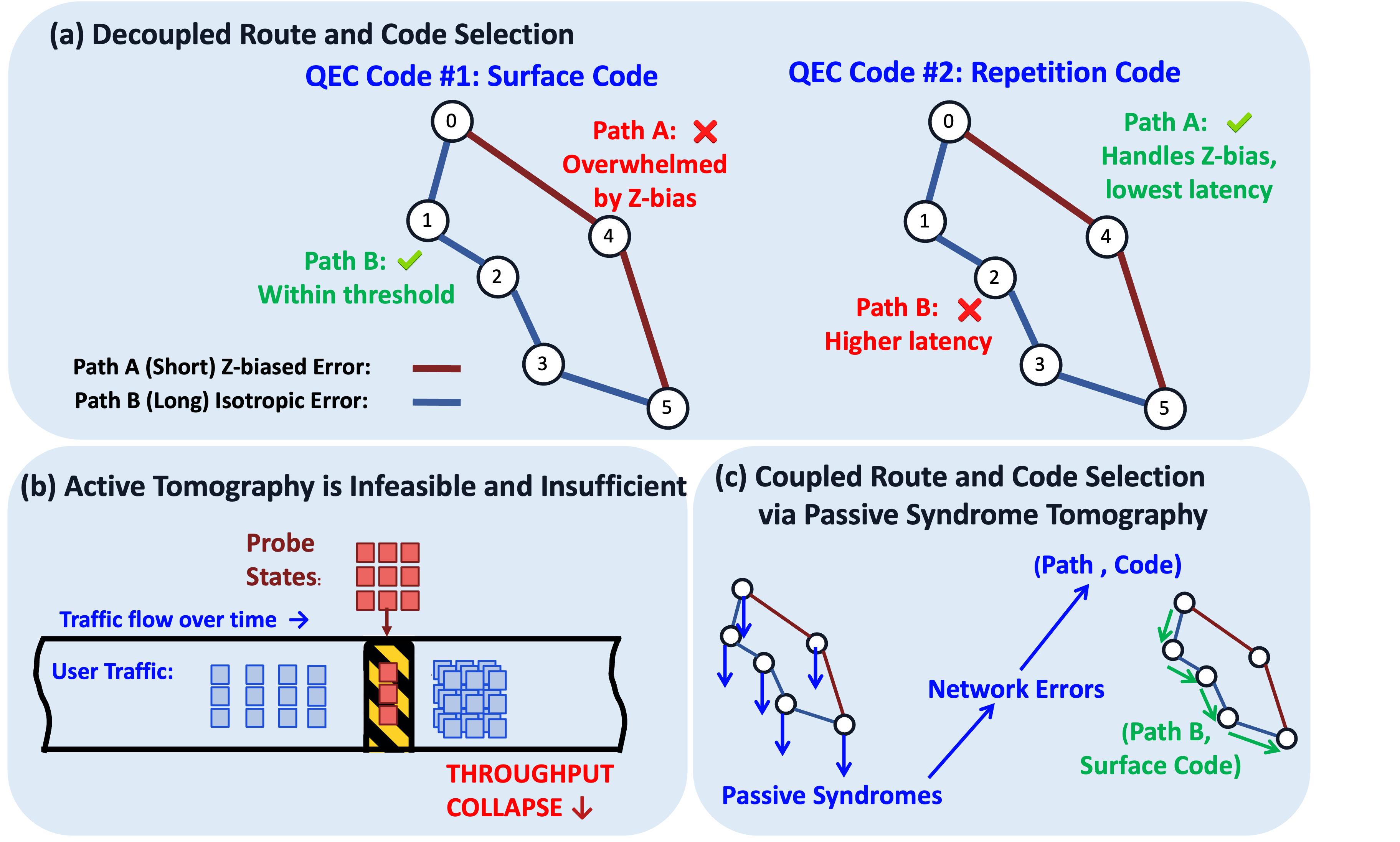}
    \caption{Motivation for coupled route and code selection, via passive syndrome probes.}
    \vspace{-0.2in}
    \label{fig:motivation}
\end{figure}

\spara{The Failure of Topology-Based Routing.}
In classical networks, ``shortest path'' is a robust proxy for performance.
In quantum networks, however, minimizing hop count or maximizing physical link fidelity does not guarantee minimal logical error rates.
Consider a scenario with two paths between a source and destination (see Fig.~\ref{fig:motivation}):
\begin{packeditemize}
    \item \textbf{Path A (Short):} One link suffers from strong $Z$-biased noise (phase flips) due to a specific hardware impairment.
    \item \textbf{Path B (Long):} All links have uniform noise.
\end{packeditemize}
A standard topology-based router will select \textbf{Path A} to minimize resource usage.
However, if the application uses a standard Surface Code (which has symmetric thresholds for $X$ and $Z$ errors), the strong $Z$-bias on Path A may overwhelm the decoder, resulting in a logical failure.
Conversely, \textbf{Path B}, despite being longer and consuming more raw entanglement, might have a noise profile that stays well within the code's correction threshold, yielding a higher goodput.
Crucially, the ``correct'' decision flips if the application changes its code: if the user switches to a Repetition Code (which is immune to $Z$ errors), Path A becomes optimal.
\textbf{The Takeaway:} Routing cannot be divorced from coding. An ``efficient'' route is undefined without knowing the code, and an ``optimal'' code depends on the specific noise bias of the route.
To solve this, we must (i) first, determine the detailed map of noise biases (not just scalar fidelities) across the network, and (ii) then, \emph{jointly} optimize the code and route for each transmission.

\spara{The "Observer Effect" of Active Tomography.}
The standard approach to characterize link noise profiles is \emph{active} tomography~\cite{blume2017demonstration, chuang1997prescription}: taking links offline to run dedicated probe circuits (e.g., Gate Set Tomography). For a production network, this is prohibitive for several reasons: (i) \emph{Throughput Collapse:} Probing requires halting user traffic. High-precision tomography can require thousands of shots, resulting in significant service interruptions. (ii) \emph{Context Mismatch:} Probing an idle network fails to capture the correlated or load-dependent error patterns present during active traffic.
Furthermore, active probes characterize links in isolation, forcing the control plane to rely on idealized composition models that fail to accurately predict end-to-end path fidelity.

\para{The Solution: Passive Syndrome Fingerprinting.} We need a signal that is (1) rich enough to distinguish error types (biases) and (2) cheap enough to collect continuously. Standard passive metrics like End-to-End Fidelity or Success Rates are insufficient; knowing a link has 90\% fidelity tells us \emph{that} it is noisy, but not \emph{how} (e.g., is it dominant X-noise or Z-noise?), leaving the router blind to the optimal code choice. Syndromes fill this gap. Generated automatically by QEC decoders during normal operation, syndromes are parity measurements that flag stabilizer violations in the quantum state. While a single syndrome is ambiguous, the \emph{statistical distribution} of syndromes collected over a window acts as a unique fingerprint of the path's physical noise process. \scope leverages this insight: by aggregating these "free" application-layer signals,\footnote{We assume the destination performs syndrome measurement and correction by default. This is required to: (a) purify the received state before application usage, and (b) prevent physical error accumulation from exceeding the code's correction threshold (i.e., the code distance).} 
we can infer the detailed physical error map required for joint routing-coding optimization without ever halting the network for probing.

\section{System Architecture}
\label{sec:system}

\begin{figure}
    \centering
    \includegraphics[width=0.5\textwidth]{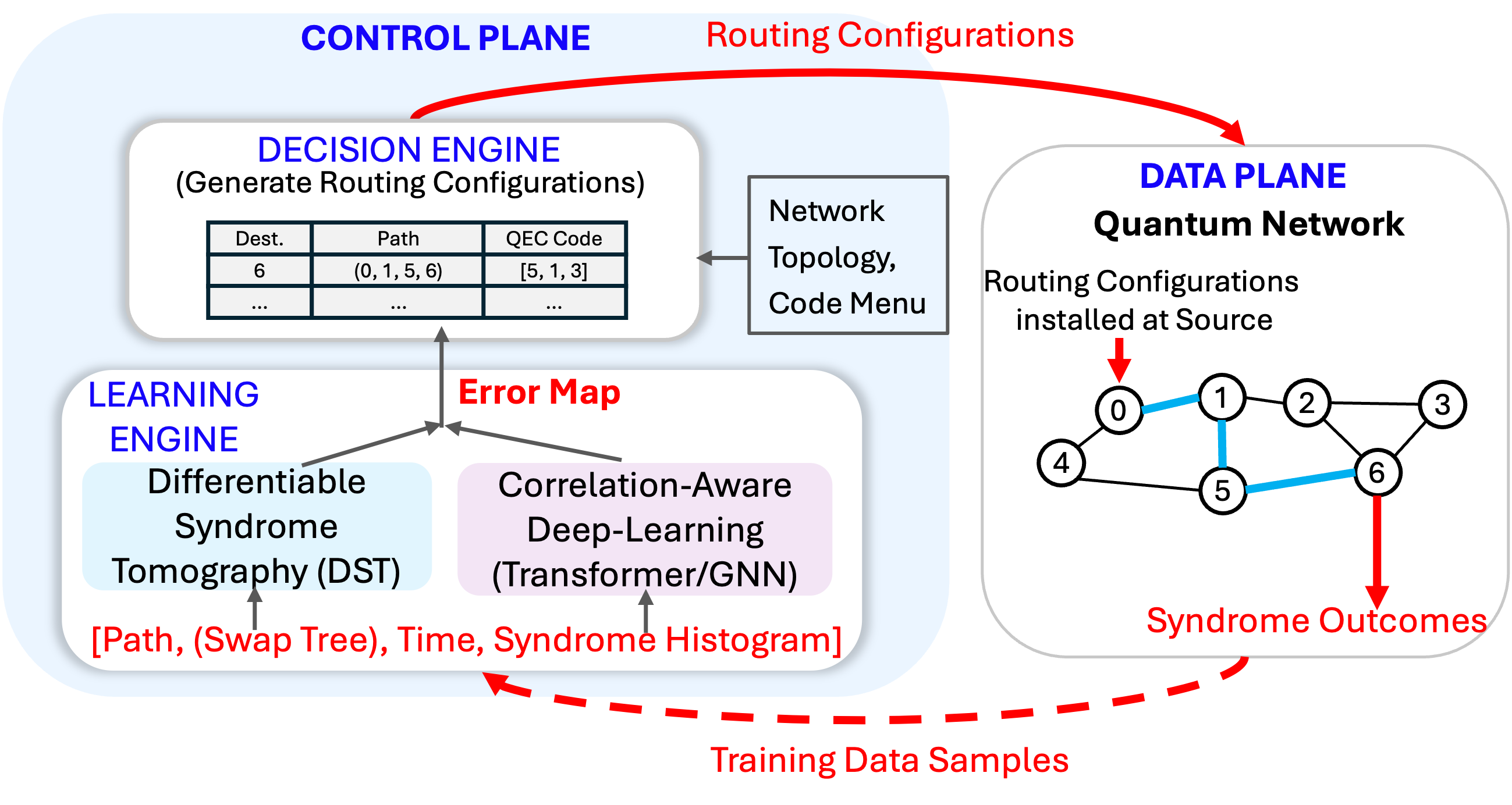}
    \caption{\small \scope System Overview. The architecture follows a closed-loop Software-Defined Networking (SDN) pattern.
    The Control Plane hosts the intelligence, comprising the \emph{Learning Engine} (which infers the Error Map $\hat{\Theta}$ from passive syndromes) and the \textit{Decision Engine} (which computes optimized Route-Code plans). The Data Plane executes these plans, transmitting qubits, generating remote entanglement, and exporting decoding-syndrome outcomes as feedback.}
    \vspace*{-0.25in}
    \label{fig:overview}
\end{figure}

We design \scope as a logically centralized control plane that makes QEC-protected networking practical at scale. As illustrated in Figure~\ref{fig:overview}, the system operates a continuous feedback cycle: it aggregates passive error syndromes from the  network, infers the underlying noise environment, and dynamically reconfigures routing tables to minimize LER.

By decoupling the complex decision logic from the real-time transmission hardware, \scope establishes a robust Software-Defined Quantum Network (SDQN) architecture comprising two primary modules:

\para{Module 1: The Tomography Engine (The "Eye").}
This module is responsible for visibility. It continuously ingests a stream of \emph{passive feedback}—specifically, the syndrome outcomes from end-point decoders and associated routing metadata (e.g., path IDs or timestamps).
Unlike active tomography, 
this engine operates in the background on live traffic.
It employs the hybrid learning architecture detailed in \S\ref{sec:indep}-\ref{sec:gen}:
\begin{packeditemize}
    \item \textbf{\dstt Engine:} Used to efficiently estimate network error parameters assuming independent link errors.
    \item \textbf{Correlation-Aware Engine (Deep Learning):} In conjunction with \dstt, it predicts network errors in the more general dependent-error model.
\end{packeditemize}
The output is a dynamic \emph{Network Error Map} $\hat{\Theta}$ that provides a unified, high-fidelity view of the network's reliability state.

\para{Module 2: The Decision Engine (The "Brain").}
This module transforms visibility into action.
Leveraging the inferred Error Map, the network topology, and a pre-defined \emph{Code Menu}, this engine executes the joint optimization algorithms described in \S\ref{sec:selection_of_route_code}. In essence, for every source-destination pair, it identifies the specific tuple—\texttt{(Route, Swap Tree, QEC Code)}—that minimizes the predicted end-to-end logical error rate.
Continuously updated (see \S\ref{sec:gen}) \emph{routing configurations} are pushed to network nodes to optimize routing.


\para{Control-Data Plane Separation and Circuit-Switching.} \scope addresses the latency constraints of quantum hardware by strictly separating policy generation from execution. The {\em Control Plane} (hosted on a classical cluster) handles the computationally intensive tasks of learning  
 $\hat{\Theta}$ 
  and solving the joint routing-coding optimization. Crucially, these complex plans are distributed \emph{proactively} and \emph{asynchronously} to source nodes, decoupling the optimization time from the request critical path. The {\em Data Plane} remains lightweight and follows a \emph{source-routing} paradigm. When a request arrives, the source node retrieves the pre-calculated strategy (e.g., specific path and code) from a local table—a microsecond-scale lookup—and initiates a \emph{connection-oriented setup phase}. Using a standard reservation protocol (or SDN flow modification)~\cite{aguado2019engineering}, the source signals downstream nodes to lock optical switches or reserve entanglement resources along the route. This ensures that the physical circuit is established before any quantum information is transmitted, eliminating the need for hop-by-hop routing logic on flying qubits. 



\para{Congestion Control.}
While explicit rate adaptation is handled by transport-layer protocols, \scope provides intrinsic congestion avoidance at the routing layer.
This is because, in quantum networks, congestion manifests physically as \emph{decoherence}: qubits waiting for busy links or swap resources must sit in memory, accumulating errors—which are naturally detected by syndrome measurements over qubits using that link/path.
In essence, a congested link or node naturally exhibits a spiked error profile.
Consequently, the decision engine automatically downgrades the quality score of congested paths and steers traffic toward less utilized routes.
This unifies error correction and load balancing: the system avoids congestion not by tracking queue lengths, but by minimizing the noise penalty associated with waiting.

\section{Estimation of Network Errors in Independent-Error Model}
\label{sec:indep}

This section details the mechanism for estimating network error parameters from observed syndromes. \textbf{For clarity of presentation}, we adopt the following base assumptions (relaxed subsequently in \S\ref{sec:dep} and \S\ref{sec:gen}):

\begin{packedenumerate}
    \item \emph{Independent Errors:} Errors on distinct links are statistically independent. (Correlated errors addressed in \S\ref{sec:dep}).
    \item \emph{Channel Errors Only:} We focus initially on Pauli channel errors. (Decoherence errors incorporated in \S\ref{sec:gen}).
    \item \emph{Direct Transmission:} We assume direct qubit transmission. (Teleportation-based transport detailed in \S\ref{sec:gen}).
\end{packedenumerate}

\subsection{Training Data and Error Estimation}

\para{Network Error Parameters.}
Let $G=(V,E)$ denote the quantum network topology.
We adopt the standard per-edge error model for direct-transmission networks~\cite{nielsen2000quantum},
characterizing each edge $e\in E$ as a Pauli channel parameterized by $\theta_e=(p_e^{X},p_e^{Y},p_e^{Z})$.
Here, $p_e^{X}$, $p_e^{Y}$, and $p_e^{Z}$ represent the probabilities of the discrete Pauli errors $X$ (bit-flip), $Y$ (bit-phase-flip), and $Z$ (phase-flip), respectively.
The probability of identity (no error) is given by $p_e^{I}=1-(p_e^{X}+p_e^{Y}+p_e^{Z})$.
We denote the set of network error parameters as $\Theta=\{\theta_e\}_{e\in E}$.

\para{Syndrome Observations ($\mathbf{h}^{(i)}_{\mathrm{obs}}$).}
Consider a specific path $\mathcal{P}_i = (e_{i1}, e_{i2}, \ldots, e_{iL_i})$ that serves $M_i$ qubit transmissions.
Upon receiving each qubit $m$, the destination performs QEC decoding, generating a syndrome outcome $s^{(i,m)} \in \{0,\ldots,2^r-1\}$, where $r$ denotes the number of syndrome bits defined by the code.
\blue{Individual syndromes are stochastic; their distribution is quasi-static over short epochs but drifts with environmental conditions, calibration, crosstalk, and load/decoherence effects.} 
We aggregate these raw measurements into a normalized syndrome histogram 
$\mathbf{h}^{(i)}_{\mathrm{obs}}[s]
= \frac{1}{M_i}\sum_{m=1}^{M_i}\mathbb{I}\{s^{(i,m)} = s\},$ for $s \in \{0,1,\ldots,2^r-1\}.$
Each entry in our training set is thus a tuple $(\mathcal{P}_i, \mathbf{h}^{(i)}_{\mathrm{obs}})$.
The histogram dimension is code-dependent; for example, the 
$\llbracket 5,1,3 \rrbracket$ code ($r=4$) yields an observation vector in $\mathbb{R}^{16}$.

\para{Error Estimation Problem.}
Given the dataset $\{(\mathcal{P}_i, \mathbf{h}^{(i)}_{\mathrm{obs}})\}_{i=1}^{N}$, our objective is to estimate the global network error parameters $\Theta$ that minimizes the statistical divergence between the observed histograms and the theoretical distributions predicted by our model.
For a specific path $\mathcal{P}_i$, let $\Theta_i \subseteq \Theta$ denote the subset of edge parameters involved in the transmission.
Let $\hpredi(\Theta_i)$ represent the end-to-end syndrome histogram predicted by these parameters.
We solve for the optimal parameters $\Theta^*$ by minimizing the cumulative Kullback--Leibler (KL) divergence over all observed paths:
\vspace{-0.1in}
\begin{equation}
\Theta^* = \mathop{\arg\min}_{\Theta} \sum_{i=1}^{N}
D_{\mathrm{KL}}\!\left( \mathbf{h}^{(i)}_{\mathrm{obs}} \,\middle\|\, \hpredi(\Theta_i) \right).
\vspace{-0.1in}
\label{eq:indep_estimation}
\end{equation}
The KL divergence is our loss function, quantifying the information loss when $\hpredi$ approximates $\mathbf{h}^{(i)}_{\mathrm{obs}}$:
\vspace{-0.1in}
\[
D_{\mathrm{KL}}\!\left( \mathbf{h}^{(i)}_{\mathrm{obs}} \,\middle\|\, \hpredi \right)
= \sum_{s=0}^{2^r-1} \mathbf{h}^{(i)}_{\mathrm{obs}}[s]\,
\log\left(\frac{\mathbf{h}^{(i)}_{\mathrm{obs}}[s]}{\hpredi[s]}\right).
\]
\vspace{-0.1in}



\para{Why Not Black-Box Learning?}
Since we only observe end-to-end syndromes—and not the ground-truth physical error rates—it might seem simpler to treat the network as a black box and learn a direct mapping from paths to syndrome histograms.
We reject this approach because it is inefficient and brittle.
A black-box model treats paths as \emph{independent entities}, failing to leverage the fact that different paths share underlying edges (and thus share error characteristics).
Consequently, it cannot generalize to unobserved routes or enforce physical constraints, such as probability normalization.
By explicitly estimating per-edge parameters, our framework enables the controller to "compose" known links to predict the fidelity of any potential path, enabling scalable, predictive routing. 


\subsection{Differentiable Syndrome Tomography}
\label{sec:dst}

\para{Method Overview.}
We introduce \textit{Differentiable Syndrome Tomography} (\dstt), a framework that estimates network error parameters by modeling quantum error propagation as a physics-informed, differentiable computational graph.
\dstt takes the data samples $(\mathcal{P}_i, \mathbf{h}^{(i)}_{\mathrm{obs}})$ and estimates $\Theta$ so that the implied end-to-end syndrome histograms closely match the observed histograms, effectively ``inverting'' the error accumulation process.
\dstt operates purely on passive, end-to-end syndrome data, and 
avoids intermediate syndrome measurements, which are not only difficult to implement but also introduce additional gate errors. Crucially, by formulating the Pauli error laws as a differentiable function linking $\Theta$ to the output histograms, \dstt ensures that gradients flow back to the shared edge parameters, constraining each $\theta_e$ using evidence from every path that traverses it.

\begin{figure}
    \centering
    \includegraphics[width=0.4\textwidth]{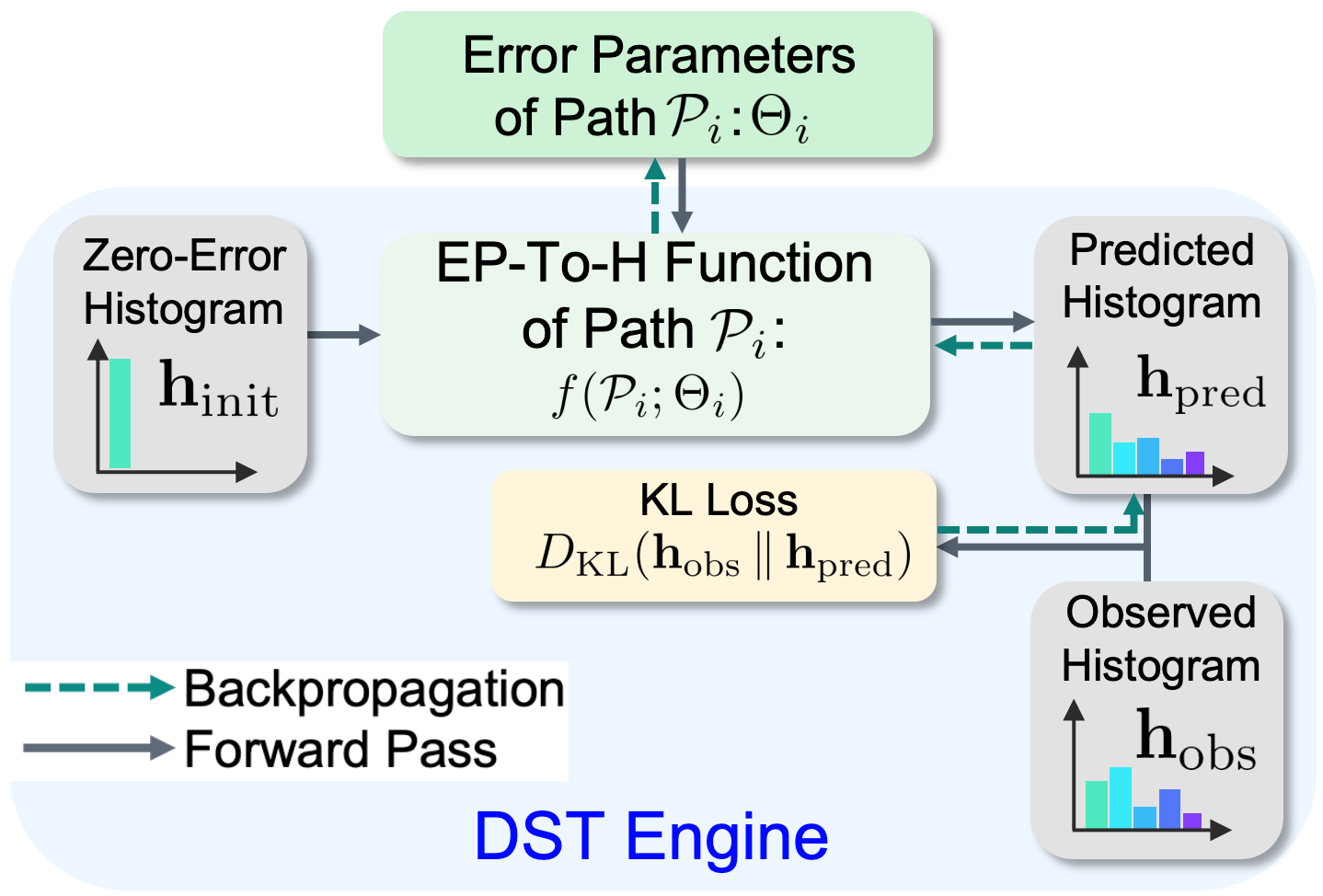}
    \caption{\dstt Engine Architecture}
    \vspace{-0.2in}
    \label{fig:dstt_arch}
\end{figure}

\para{Errors-to-Histogram (\eh) Forward Mapping $f(.)$.}
To facilitate error estimation, we model the relationship between physical error parameters and end-to-end syndrome outcomes as a ``forward'' mapping $f(\mathcal{P}_i;\Theta_i)$.
Formally, for a path $\mathcal{P}_i$, we define the predicted histogram as:
$
f(\mathcal{P}_i;\Theta_i)=\hpredi,
$
where $\Theta_i$ denotes the learnable parameters associated with the edges in $\mathcal{P}_i$.
To enable gradient-based optimization, we construct $f(\cdot)$ as a differentiable computational graph composed of three steps:

\softpara{1. Basis Construction.}
We implement the mapping via matrix algebra.
Taking the $\llbracket5,1,3\rrbracket$ code as a concrete example, the syndrome space has size $2^r=16$.
We first enumerate all $4^5=1024$ possible physical Pauli error configurations on the 5-qubit link:
$
\mathcal{C}=\{(Q_1,\ldots,Q_5)\mid Q_k\in\{I,X,Y,Z\}\}.
$
A specific error configuration $c \in \mathcal{C}$ acts as a deterministic update to the syndrome state, shifting an input syndrome $j$ to $j \oplus s(c)$, where $s(c)$ is the syndrome of error $c$.
This is represented as a pre-computed permutation matrix $\mathbf{B}_c$: 
where $(\mathbf{B}_c)_{ij}= 1$ if $i=j\oplus s(c)$ and is 0 otherwise.


\softpara{2. Edge Transition Matrix.}
For a specific network edge $e$, the syndrome transition is a probabilistic mixture of these basis permutations, determined by the edge's physical error rates $\theta_e$.
We synthesize the edge transition matrix $\mathbf{M}_e(\theta_e)$ as:
$
\mathbf{M}_e(\theta_e)=\sum_{c\in\mathcal{C}} w_e(c)\,\mathbf{B}_c,
\quad \text{where }
w_e(c)=\prod_{k=1}^{5} p^{Q_k}_e.
$
Here, the weight $w_e(c)$ is the probability of configuration $c$ occurring, derived from the per-edge Pauli probabilities (e.g., using $p^X_e$ if $Q_k=X$).

\softpara{3. Path Propagation.}
Finally, to compute the end-to-end histogram for path $\mathcal{P}_i=(e_{i1},\ldots,e_{iL_i})$, we propagate an initial "no-error" distribution vector $\mathbf{v}_0=[1,0,\ldots,0]^\top$ through the sequence of edge matrices to yield the mapping as:
\[
f(\mathcal{P}_i;\Theta_i)
\triangleq \hpredi 
= \left(\prod_{k=1}^{L_i}\mathbf{M}_{e_{ik}}(\theta_{e_{ik}})\right)\mathbf{v}_0.
\]
This formulation yields a closed-form, differentiable expression for the predicted histogram.
Since each $\mathbf{M}_e$ is a smooth function of $\theta_e$, the mapping $f(\mathcal{P}_i;\Theta_i)$ is differentiable with respect to $\Theta$; we 
can thus compute gradients of the loss function with respect to $\Theta$ via standard backpropagation.

\para{Optimization Procedure.}
Substituting the differentiable \eh mapping into our objective, we define the loss function $\mathcal{L}(\Theta)$ as the cumulative KL divergence over the dataset:
\begin{equation}
\mathcal{L}(\Theta) = \sum_{i=1}^{N} \sum_{s=0}^{2^r-1}
\mathbf{h}^{(i)}_{\mathrm{obs}}[s] \,
\log\left(\frac{\mathbf{h}^{(i)}_{\mathrm{obs}}[s]}
{\hpredi[s] + \epsilon}\right),
\label{eq:loss_function}
\end{equation}
where $\epsilon$ is a small smoothing constant (e.g., $10^{-9}$) added to ensure numerical stability.
We minimize this loss subject to the physical constraint that error probabilities on each edge must form a valid distribution (sum to $\le 1$).

We solve this constrained optimization using \textit{Projected Gradient Descent} (PGD).
Starting from an initialization $\Theta^{(0)}$, we iteratively update the parameters:
\[
\Theta^{(t+1)} \leftarrow \Pi_{\mathcal{S}}\left( \Theta^{(t)} - \eta_t \nabla_{\Theta}\mathcal{L}(\Theta^{(t)}) \right),
\]
where $\eta_t$ is the learning rate, $\nabla_{\Theta}\mathcal{L}$ is the gradient computed via backpropagation, and $\Pi_{\mathcal{S}}$ denotes the projection onto the valid probability simplex.

\para{Convergence and Generalization.}
Even in regimes where exact parameter recovery is theoretically underdetermined, our estimation converges to an error profile that is \textit{statistically equivalent} to the observed histograms.
Crucially, the physical constraint that a single set of parameters $\theta_e$ is shared across all paths traversing edge $e$ introduces strong \emph{cross-path coupling}.
This coupling significantly shrinks the feasible solution space, encouraging convergence toward the true physical parameters rather than path-specific overfitting.
Finally, because our model is compositional, the estimated $\Theta$ generalizes beyond the training data: it allows the controller to predict the fidelity of \emph{unseen} paths using \emph{seen} edges.


\section{Learning Path-Dependent Errors}
\label{sec:dep}

The independent-error model in \S\ref{sec:indep} assumes that the Pauli error parameters on an edge are independent hardware properties.
However, in physical quantum networks, the \textit{effective} Pauli error rate often depends on the qubit's prior state and trajectory.
Here, we relax the independence assumption to address this \textit{path-dependent} nature of quantum noise.

\para{Motivation: Physical Sources of Dependency.}
Most network tomography methods assume link errors are statistically independent.
In quantum networks, however, an edge's effective noise is inherently history-dependent, arising primarily from \textit{coherent errors} and \textit{switching context}.
Although we represent errors as stochastic Pauli flips for tractable estimation, physical noise is typically coherent (e.g., systematic unitary rotations).
Such noise accumulates vectorially, meaning it can constructively amplify or destructively interfere depending on the sequence of operations along the path~\cite{ExampleCoherent}.
Consequently, the \textit{effective} Pauli error probability manifested on a link depends on the qubit's prior trajectory.
Additionally, active optical components (switches) often exhibit configuration-dependent crosstalk, further coupling the error rate to the specific routing path~\cite{SwitchPaper}.
Crucially, this path dependence is even more pronounced in teleportation-based networks (addressed in \S\ref{sec:gen}).
In that setting, probabilistic entanglement distribution forces qubits to wait in memory while remote links are generated.
The resulting decoherence error is strictly determined by the \textit{temporal history} of link generation along the path~\cite{Hartmann2007Role, Goodenough2025SwapASAP}.

Thus, regardless of whether the mechanism is spatial (coherent interference) or temporal (memory decoherence), accurate estimation requires treating the error channel not as a static constant, but as a dynamic variable conditioned on the path context.

\para{Formalization.}
Developing a unified analytical model that captures all sources of dependence—ranging from coherent interference to device-specific memory effects—is intractable for real-world networks.
Instead of attempting to model specific physical mechanisms, we abstract the path dependence into a generic recursive form:
\[
\theta_{e_k} = g(e_k \mid z_k), \qquad
z_k = \psi(e_1,\ldots,e_{k-1}),
\]
where $z_k$ denotes a hidden \textit{context state} determined by the sequence of preceding edges.
Thus, the error parameters of the $k$-th edge are conditioned on the upstream path history encapsulated in $z_k$.
Prior theoretical models can be viewed as special cases of this template, differing primarily in their structural assumptions for the state transition function $\psi(\cdot)$.
Since the true form of $\psi$ is complex and hardware-specific, we do not define it analytically; instead, we introduce a machine-learning approach to learn these dependencies in a model-agnostic manner directly from data.

\para{Challenges.}
Implementing this context-dependent model introduces two fundamental challenges:

\begin{packedenumerate}
\item \underline{Information Loss in Static Estimation.}
In a path-dependent regime, an edge's error profile varies dynamically.
By contrast, independent-error techniques like \dstt enforce a single, global parameter set for each edge.
Consequently, when applied to path-dependent noise, such estimators inevitably converge to a \textit{marginal average} of the edge's behavior across observed traffic.
While this average might describe general network health, it fails to capture critical context-specific deviations (e.g., ``Edge A is good, unless it follows Edge B'').
This renders static estimation unreliable for fidelity-aware routing, which requires precise predictions for specific path combinations rather than just average link metrics.

    \item \underline{Unobservable Latent Dynamics.}
    The context state $z_k$ is latent—it cannot be measured directly.
    Furthermore, as noted above, the transition function $\psi$ is complex and hardware-specific; we cannot rely on predefined analytical models (like simple memory decay) that may not hold for a given device.
    Therefore, we face the challenge of learning a high-dimensional mapping from path history to error parameters in a \textit{function-agnostic} manner, using only sparse end-to-end syndrome data and without ground-truth labels for the intermediate states.
\end{packedenumerate}

\begin{figure}
    \centering
    \includegraphics[width=0.9\linewidth]{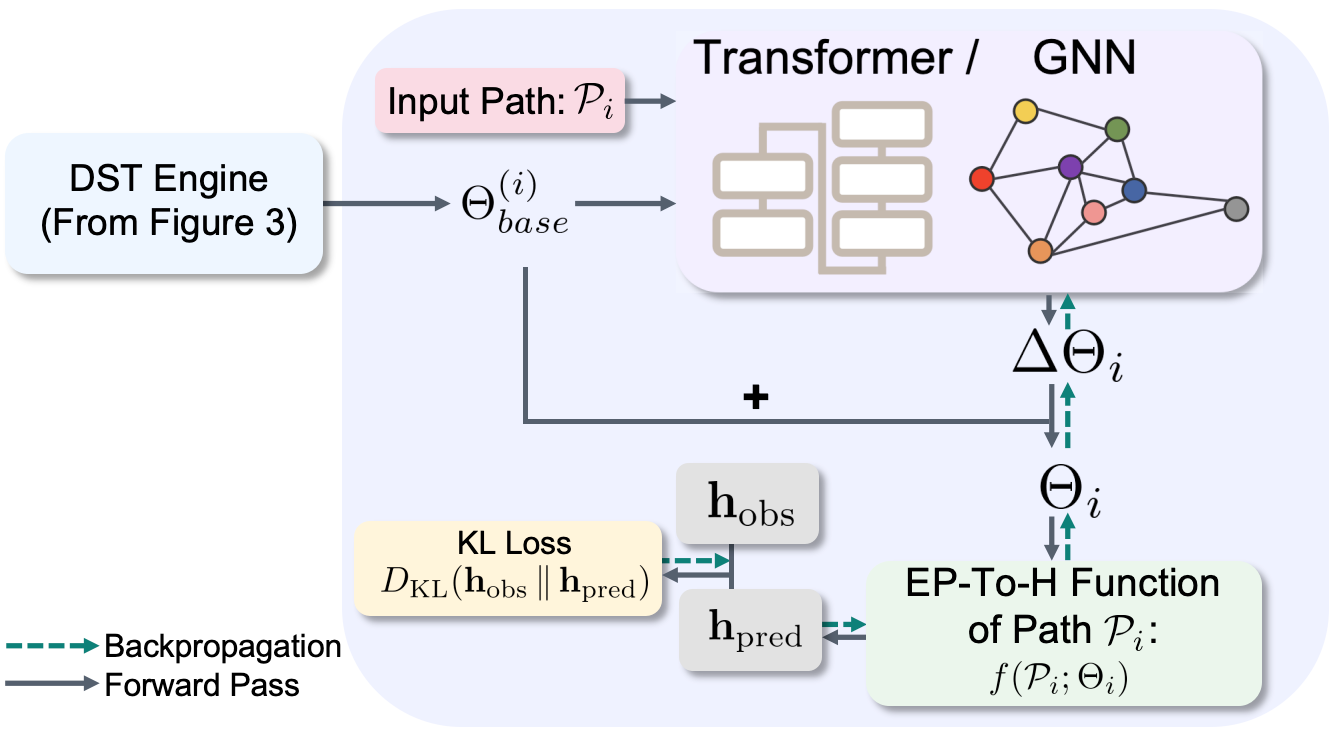}
    \caption{\small Correlation-Aware Learning Engine}
    \vspace{-0.2in}
    \label{fig:ce_arch}
\end{figure}
\subsection{Correlation-Aware Learning Engine}

\para{Design Goals.}
To address the challenges of path-dependent noise, we seek a tomography framework that satisfies three key properties:
\begin{packedenumerate}
    \item \emph{Context Awareness:} The model must produce \emph{dynamic} edge parameters, allowing the effective error rate of a link to vary based on the specific path history preceding it.
    \item \emph{Mechanism Agnosticism:} It should be data-driven rather than model-based; instead of relying on handcrafted analytical assumptions for the dependency function $\psi$, it must learn the latent physics directly from observed traffic.
    \item \emph{Physical Consistency:} It should remain physically grounded by integrating with the analytic error-propagation model derived in \S\ref{sec:indep}. This hybrid approach ensures data efficiency and guarantees that the inferred parameters always form a valid probability distribution.
\end{packedenumerate}
Motivated by the above, we propose a two-stage hybrid architecture that combines the physical interpretability of \dstt with the expressive power of deep learning. See Fig.~\ref{fig:ce_arch}.

\para{Stage 1: Static Baseline Estimation.}
First, we run the physics-based \dstt method (described in \S\ref{sec:indep}) on the full dataset (syndrome histograms) to learn a path-independent parameter table $\Theta_{\text{base}} = \{\theta^{\text{base}}_e\}_{e\in E}$.
As discussed in \S\ref{sec:dep}, since \dstt enforces a single parameter set per edge, the resulting estimate $\theta^{\text{base}}_e$ converges to a {\it traffic-weighted marginal average} of the edge's behavior across all observed contexts. This provides a stable initialization for Stage~2.

\para{Stage 2: Contextual Refinement.}
In the second stage, we freeze $\Theta_{\text{base}}$ and train a deep neural network, the \emph{Edge Adjustment Network (EAN)}, to capture residual dependencies.
For a given path $\mathcal{P}_i$, the EAN predicts a context-specific adjustment vector $\Delta\theta^{(i)}_{e}$ for every edge $e \in \mathcal{P}_i$.
The effective edge parameters are then modeled as:
\[
\theta^{(i)}_{e} = \text{clip}\left( \theta^{\text{base}}_{e} + \Delta\theta^{(i)}_{e} \right),
\]
where the clip operation ensures that parameters remain within $[0, 1]$. These effective parameters are fed into the  \eh mapping to compute the predicted histogram $\hpredi$.

\softpara{Optimization Objective.}
We train the EAN parameters $\phi$ by minimizing the divergence between the predicted and observed histograms, for a fixed $\Theta_{\text{base}}$, regularized by an $\ell_2$ penalty to minimize deviations from the physics baseline:
\[
\phi^{\star} = \arg\min_{\phi} \sum_{i=1}^{N} \left( D_{\mathrm{KL}}\!\left( \mathbf{h}^{(i)}_{\mathrm{obs}} \,\middle\|\, \hpredi(\phi) \right) + \lambda \sum_{e \in \mathcal{P}_i} \|\Delta\theta^{(i)}_{e}\|_2^2 \right).
\]
This objective ensures that the model introduces context dependence only when the static baseline is insufficient to explain the data.

\para{EAN Implementations: Transformer vs. GNN.}
The EAN is a plug-in module that can be instantiated with different neural architectures depending on the dominant dependency source. We consider two variants:

\begin{packeditemize}
    \item \textbf{Sequential Dependence (Transformer):}
    To capture history-dependent effects (e.g., decoherence from accumulated delays), we implement the EAN as a \emph{Transformer}~\cite{vaswani2017attention}.
    It treats a path $\mathcal{P}_i$ as an ordered sequence of tokens and uses causal masked self-attention to predict adjustments.
    The causal mask ensures that the adjustment for $e_k$ depends strictly on the upstream history $\{e_1, \dots, e_{k-1}\}$, matching the physical causality of quantum memory errors.

    \item \textbf{Structural Dependence (Graph Neural Network):}
    To capture topology-induced effects (e.g., crosstalk from neighboring links), we implement the EAN as a \emph{Graph Neural Network (GNN)}~\cite{scarselli2008graph}.
    This variant operates on the full network graph $G=(V, E)$.
    It performs message passing to aggregate information from the local neighborhood of each edge in the path, allowing the adjustment $\Delta\theta$ to reflect structural interference patterns that a purely sequential model might miss.
\end{packeditemize}

\cblue

\section{Extensions to General Network Settings}
\label{sec:gen}

We now extend the \scope framework beyond the baseline model to capture several practical realities of QEC-enabled quantum networks:
\begin{packedenumerate}
    \item \emph{Decoherence and Storage:} We incorporate time-dependent errors caused by qubit idling and memory buffering during transmission.
    
    \item \emph{Teleportation-Based Transport:} We extend the forward model to entanglement-swapping chains, accounting for swap-operation errors and waiting-time-induced decoherence.
    
    
    \item \emph{Heterogeneous QEC Codes:} We support traffic encoded with different QEC codes by sharing the same physical error model $\Theta$ while using code-specific syndrome mappings, enabling joint training across code families.
    
    \item \emph{Cold Starts and Topology Changes:} We describe how \scope bootstraps with limited initial telemetry and adapts to link/node failures, replacements, and additions without full retraining.
    
    \item \emph{Continuous Adaptation:} We show how \scope maintains model freshness under non-stationary noise by incrementally updating its estimates from newly collected syndrome observations.
\end{packedenumerate}
\cb

\para{Decoherence Error.}
Our primary formulation assumes errors are purely spatial (link-specific). 
In reality, qubits buffering in quantum memories accumulate \emph{decoherence} errors (primarily $Z$-bias) proportional to their storage duration. 
To capture this, we leverage the \emph{total transmission time} ($\tau$) extracted from classical control plane timestamps—a reasonable assumption given that quantum link protocols require nanosecond-level synchronization~\cite{dahlberg2019link, alshowkan2021advanced}.
The data sample is now the tuple $(\mathcal{P}_i, \tau, \mathbf{h}^{(i)}_{\mathrm{obs}})$, where $\tau$ is the mean latency of transmissions over $\mathcal{P}_i$ during a stability epoch---a short window wherein network queue depths and resulting decoherence effects remain quasi-stationary.
\blue{We note that \scope does not require cross-node clock synchronization; timestamps are used only to estimate the decoherence-induced contribution to path error.}

\softpara{Forward Model Adaptation.} 
The forward function $f(\cdot)$ extends to include learnable node parameters $\theta_{v} = \{T_{1,v}, T_{2,v}\}$ for each node $v$. 
Assuming the total latency $\tau$ is uniformly distributed across the $L$ nodes in the path, the error probability contributions for node $v$ are
modeled as standard Pauli-twirled channels~\cite{nielsen2000quantum}:
$p_{x,v}(\tau) = p_{y,v}(\tau) \approx 1/4(1 - e^{-(\frac{\tau}{L \cdot T_{1,v}})}), p_{z,v}(\tau) \approx 1/2(1 - e^{-( \frac{\tau}{L \cdot T_{2,v}})})$.
Here, $T_{1,v}$ governs energy relaxation (contributing to $X, Y$ errors) and $T_{2,v}$ governs dephasing (dominating $Z$ errors).
The total path error profile is then the sum of the static link errors (spatial) and these dynamic node errors (temporal). 
By training on these temporal batches, SCOPE implicitly learns the coherence times ($T_{1,v}, T_{2,v}$) of the network nodes, allowing the router to predict contributions to LERs arising from decoherence delays. 

\para{Teleportation-Based Transport.}
Thus far, we have assumed a \textit{direct-transmission} model where qubits physically traverse a sequence of edges.
However, scalable quantum networks typically employ \textit{teleportation}, in which end-to-end connectivity is achieved by generating elementary entanglement on links and performing swapping (BSM) operations at intermediate nodes.
This introduces distinct noise mechanisms—specifically, swapping errors and memory decoherence—that require a more sophisticated forward mapping.

\softpara{Teleportation Mapping Function $f_{\mathrm{tele}}$.}
In a teleportation network, end-to-end noise is not merely the sum of edge errors; it is a composite of link-entanglement quality, the fidelity of swapping operations, and decoherence induced by waiting times.
To capture this, we extend the prior \eh mapping to include two sets of learnable variables:
\begin{itemize}
    \item $\Theta_{\mathrm{edge}} = \{\theta_e\}$: Parameters for entanglement generation on links.
\item $\Theta_{\mathrm{node}} = \{\theta_v\}$: Parameters characterizing storage coherence ($T_{1,v}, T_{2,v}$) and operational Pauli errors ($p_{x,v}, p_{y,v}, p_{z,v}$) arising from swapping (BSM) and frame updates.
\end{itemize}
The syndrome-histogram prediction model then becomes:
\[
\hpredi
\;=\;
f_{\mathrm{tele}}\!\big((\mathcal{P}_i, \mathcal{T}^{(i)}); \Theta_{\mathrm{edge}}, \Theta_{\mathrm{node}}\big),
\]
where $\mathcal{T}^{(i)}$ represents the \textbf{Swap Tree}—the specific hierarchical order in which links were merged to form the end-to-end connection—\textbf{augmented with the waiting times} incurred at each node.
We note that these waiting times are readily available to the network controller, as they can be derived directly from the timestamps of the classical Pauli correction frames generated during the swapping process.

\softpara{Dataset.}
Each data sample is thus $((\mathcal{P}_{i}, \mathcal{T}^{(i)}), \mathbf{h}^{(i)}_{\mathrm{obs}})$, while the supervision signal remains the same: $\mathbf{h}^{(i)}_{\mathrm{obs}}$.

\softpara{Independent-Error Model.}
In an \emph{independent} error model, each node is assigned a single, static parameter $\theta_v$. This implicitly assumes that the node's performance is constant, thereby marginalizing over variations in swap errors and waiting times across requests.
To estimate the edge and node error parameters, we minimize the divergence between the observed and predicted histograms by leveraging the differentiability of the swap logic.
The extended \eh function computes the end-to-end channel by composing edge and node error over $\mathcal{P}_i$.
Gradients are backpropagated through this composition to jointly update the static parameters:
\[
\min_{\Theta_{\mathrm{edge}},\Theta_{\mathrm{node}}}
\sum_{i=1}^{N} D_{\mathrm{KL}}\!\left(
\mathbf{h}^{(i)}_{\mathrm{obs}} \,\middle\|\,
f_{\mathrm{tele}}\!\big((\mathcal{P}_i, \mathcal{T}^{(i)}); \Theta_{\mathrm{edge}}, \Theta_{\mathrm{node}}\big)
\right).
\]

\softpara{Dependent-Error Model.}
In the \emph{dependent} error model, we treat the link-entanglement parameters $\Theta_{\mathrm{edge}}$ as independent, as entanglement generation is typically an atomic, heralded process.
However, we recognize that the effective node parameters $\Theta_{\mathrm{node}}$ are highly context-sensitive.
Specifically, for a given swap topology $\mathcal{T}^{(i)}$, a node must store qubits for varying durations while waiting for upstream links to succeed, and the BSM operation itself may degrade depending on the incoming state history or node congestion.
Therefore, the model must learn two components: (i) the independent $\Theta_{\mathrm{edge}}$ parameters, and (ii) the context-aware $\Theta_{\mathrm{node}}$ parameters, which are driven by the specific tree structure (as it dictates both waiting times and BSM fidelity).
Our two-stage ML-Driven architecture (\S\ref{sec:dep}) handles this naturally:
\begin{packeditemize}
    \item \textbf{Stage 1:} Learns static baseline parameter values, capturing the average performance of edges and nodes.
    \item \textbf{Stage 2:} The Transformer or GNN takes the path $\mathcal{P}_i$ and the swap tree $\mathcal{T}^{(i)}$ as inputs to predict residual adjustments for the node parameters, and thus, learn the penalty induced by expected waiting times and BSM context.
\end{packeditemize}
This enables the framework to predict the exact fidelity of a candidate route and swap tree, without requiring prior knowledge of timestamps or internal node states.

\para{Unified Training across Heterogeneous Codes.}
Thus far, we have implicitly assumed that all training samples use the same error-correcting code. In practice, however, heterogeneous traffic may use different QEC codes over the same physical infrastructure.
Crucially, the physical error parameters $\Theta$---such as link entanglement fidelity and node memory noise---are properties of the \emph{hardware}, not the code.
Thus, instead of training separate models for each code type, we train a single unified model by augmenting each sample with its code descriptor $c^{(i)}$.
The forward pass then dynamically selects the specific mapping function $f_{c^{(i)}}$—derived from that code's unique parity check matrix—to compute the syndrome implication from the shared physical parameters $\Theta$:
$
\hpredi=
f_{c^{(i)}}\!(\mathcal{P}_i;\, \Theta).
$
Gradients from these diverse codes are back-propagated into the same shared parameter set $\Theta$.
This enables a powerful \emph{transfer learning} capability: the network can utilize abundant data from cheap, low-overhead codes (like small Repetition codes) to refine the physical error model, which is then used to optimize routing for expensive, high-fidelity codes (like large Surface codes) that are too costly to use for frequent probing.

\para{Cold Start and Network Dynamics.}
\blue{When \scope is first deployed, or when the learning engine has insufficient training data, the system falls back to a conservative default configuration: shortest-path routing with a default symmetric QEC code, e.g., the $\llbracket 5,1,3 \rrbracket$ code. As syndrome data accumulates, the error map is progressively refined, allowing the decision engine to transition from conservative defaults to optimized route/code plans.}

\softpara{Link and Node Failures/Additions.}
\blue{When a link or node fails, the controller removes the affected element from the feasible topology and recomputes route/code plans over the updated graph. Importantly, \scope does not discard the entire learned error map: profiles for unaffected links and nodes are preserved, and only plans that traverse the failed element are invalidated. Newly added links or nodes are initialized with conservative noise profiles. As traffic traverses these elements, their estimated parameters are refined through the standard continuous-learning loop described below.}

\para{Continuous Learning and Model Freshness.} To maintain an accurate Error Map despite hardware drift, SCOPE also employs an \emph{online learning} feedback loop. The system aggregates telemetry over sliding windows defined by a target number of syndrome shots. Rather than retraining our models from scratch—which would be too slow for real-time operations—we utilize \emph{incremental fine-tuning}. At the end of each epoch, gradients are computed on the new batch of histograms to update the existing model weights. This allows the system to track time-varying noise parameters with minimal computational lag. To prevent inefficient training on stale data, observations older than the drift timescale are discarded from the training buffer, ensuring that $\hat{\Theta}$ continuously reflects the network's latest physical state. We note that the above fine-tuning is applied only when the \emph{predicted} histogram of the path deviates from the new observed histogram by more than a threshold; this allows the system to smooth out transient noise.

\section{Route and QEC Code Selection}
\label{sec:selection_of_route_code}

This section details the SCOPE decision engine, which identifies the optimal transmission plan—comprising a route/topology $P$ and a QEC code $C$—given the inferred network error map $\hat{\Theta}$.
We formulate this as a joint optimization problem minimizing the predicted Logical Error Rate (LER).

\para{Joint Route-Code Optimization.}
The objective is to select the pair $(P, C)$ that minimizes the end-to-end logical error rate predicted by our learned model.
As motivated in Section~\ref{sec:motivation}, LER depends non-linearly on the interplay between physical noise and code properties; therefore, decoupling routing and coding is generally suboptimal.
Formally, given a source $s$, destination $d$, and a Code Menu $\mathcal{C}$, we solve:
\begin{equation}
    (P^*, C^*) = \arg\min_{P \in \mathcal{P}_{sd},\, C \in \mathcal{C}} \mathcal{D}\Big( \mathcal{M}(P, C; \hat{\Theta}) \Big),
\end{equation}
where $\mathcal{P}_{sd}$ is the set of feasible transmission structures (paths or swap trees), $\mathcal{M}$ is our learned predictor that outputs a syndrome distribution, and $\mathcal{D}$ is the decoder model that maps syndromes to an LER.
We instantiate this search for Direct Transmission and Teleportation under both independent and dependent error regimes.

\subsection{Direct Transmission}
In this mode, $P$ represents a sequence of physical links.

\para{Independent Errors: Code-Aware Shortest Path.}
When link errors are independent, the optimal path satisfies the optimal substructure property.
However, because physical error rates affect logical performance non-linearly, we cannot simply minimize the sum of physical errors.
Instead, for each candidate code $C \in \mathcal{C}$, we define a \emph{code-aware link metric} $w(e, C)$—a surrogate cost representing the reliability impact of edge $e$ on code $C$ (e.g., log-domain logical success probability).
We run Dijkstra's algorithm with these weights to identify the optimal path $P_C^*$ for each code, then select the pair $(P_C^*, C)$ that yields the minimum predicted LER.

\para{Dependent Errors: Context-Aware Routing.}
When errors are path-dependent (e.g., due to memory effects or upstream drift), standard shortest-path algorithms fail because a link's cost depends on the upstream trajectory.
To address this, we employ a context-aware Dynamic Programming (DP) scheme.
We define the state as $(v, \mathbf{h}_k)$, where $v$ is the current node and $\mathbf{h}_k$ is the history vector of the last $k$ visited nodes.
The DP value $\Phi(v, \mathbf{h}_k)$ represents the minimum LER to reach this state.
Transitions extend the path by one hop, querying the predictor $\mathcal{M}$ with the extended context:
\begin{equation}
    \Phi(v, \mathbf{h}'_k) = \min_{u \in \mathcal{N}^{-1}(v)} \left\{ \mathcal{D}\!\left( \mathcal{M}( P^*(u, \mathbf{h}_k) \oplus v, C ) \right) \right\}.
\end{equation}
This effectively unrolls the graph into a trellis of depth $k$, allowing the router to distinguish between paths that arrive at $v$ with different error-inducing histories.

\subsection{Teleportation-Based Communication}
In this mode, $P$ represents a \emph{Swap Tree} (without the waiting times, which are predicted) defining both the sequence of links and the hierarchical order of entanglement swappings.

\para{Independent Errors.}
In the independent errors model, teleportation errors depend only on the involved edges and nodes; thus, we can just select the best path using the Code-Aware Shortest Path method (as in Direct Transmission). Once the path is fixed, the swap tree topology can be determined by a latency-minimizing dynamic programming 
algorithm (e.g.,~\cite{ghaderibaneh2022efficient}) to minimize total latency.

\para{Dependent Errors: Tree Optimization.}
In the dependent regime, the hierarchical order of swaps determines waiting times and BSM contexts, directly impacting node errors.
Prior work addressed this by assuming node errors depend strictly on the local subtree structure, allowing for optimal tree-based dynamic programming (DP)~\cite{ghaderibaneh2022efficient}.
However, capturing broad dependencies—such as cross-branch resource contention or accumulated state drift—renders exact DP intractable.
Thus, we adopt a tractable two-stage strategy:
\begin{enumerate}
    \item \emph{Route Filtering:} We generate the top-$K$ candidate paths using an independent-error heuristic.
    \item \emph{Tree Construction via Interval DP:} For each candidate path and code, we optimize the swap topology using a bottom-up Interval DP described below.
\end{enumerate}
The \emph{Interval DP} constructs the optimal tree by recursively parsing the path.
Let $R[i,j]$ represent the syndrome distribution of the optimal subtree spanning $v_i \dots v_j$. To compute $R[i,j]$, we search for the split point $k$ (the final BSM location) that minimizes the decoded error $\mathcal{D}$:
\begin{equation}
    R[i,j] \leftarrow \operatorname*{argmin}_{k \in (i,j)} \Big\{ \mathcal{D}\big( \mathcal{M}(R[i,k], R[k,j], C) \big) \Big\}.
\end{equation}
Solving for the full span $R[0,L]$ yields the swap tree structure that minimizes the LER.
\emph{Note:} If the error model $\mathcal{M}$ is strictly local (dependencies limited to the immediate subtree), this Interval DP generalizes to an exact, optimal solution.

\section{Related Work}
\label{sec:related}

SCOPE sits at the intersection of network routing, quantum tomography, and error correction. While prior work has addressed these areas in isolation, SCOPE is the first to unify them into a closed-loop control plane.

\para{Quantum Routing and Resource Allocation.}
While architectural proposals for ``One-Way'' (Gen-3) networks exist~\cite{muralidharan2016optimal, munro2012quantum}, they focus primarily on point-to-point repeater physics rather than network-layer routing.
Consequently, most routing protocols focus on entanglement distribution, optimizing \emph{physical} metrics like generation rate or fidelity~\cite{pant2019routing, li2021effective, caleffi2017optimal, ghaderibaneh2022efficient}.
These approaches typically model links using scalar weights (e.g., average fidelity) combined via rigid analytical formulas.
This scalar abstraction creates two critical blind spots.
First, it masks the underlying \emph{Pauli error structure} (bias), which is decisive for logical performance. Second, analytical models fail to capture non-Markovian hardware behaviors, often assuming idealized channels that miss operational conditions like crosstalk.
{\em In contrast,} \scope explicitly infers the full Pauli error structure rather than a scalar summary.
Instead of relying on theoretical accumulation assumptions, we \emph{learn} the \emph{effective} error landscape that best explains the observed syndrome data.


\para{Syndrome-Based Estimation.}
The concept of using syndrome statistics to infer physical error rates is well-established in quantum computing theory and device characterization~\cite{fowler2014scalable, harper2020efficient,flammia2020efficient,wagner2021optimal, wagner2022pauli, wagner2023learning}.
However, these works fundamentally address a \emph{local} inverse problem: mapping the syndrome density of a single component (e.g., a specific gate or idle link) to its isolated error parameters.
In a network context, relying on such local inversion is both \emph{infeasible} and \emph{insufficient}.
It is infeasible because acquiring granular link-level syndrome data effectively mandates active probing.
It is insufficient because predicting path-level performance by analytically combining local estimates forces a return to rigid assumptions (e.g., error independence).
{\em Moreover}, deriving a global analytical inverse function for path-level errors is computationally intractable due to the exponential state space of possible error combinations and cross-link correlations.
{\em Consequently,} SCOPE abandons the attempt to derive a closed-form inverse.
Instead, we employ a learning approach that approximates this complex mapping.

\para{Quantum Tomography.}
Accurate routing demands precise noise maps.
Standard approaches such as \emph{Quantum Process Tomography}~\cite{nielsen2000quantum} or \emph{Gate Set Tomography}~\cite{blume2017demonstration} provide high-fidelity characterization, but they scale exponentially and require taking links offline. Recent \emph{compressed sensing} techniques reduce sample complexity~\cite{flammia2012quantum}, but still require injecting bandwidth-consuming probe states.
Crucially, these calibration methods suffer from \emph{context mismatch}, failing to capture load-dependent noise (e.g., crosstalk) present during heavy traffic~\cite{sarovar2020detecting}.
Recent theoretical work on \emph{Quantum Network Tomography}~\cite{de2024quantum} explicitly highlights the intractability of characterizing general network noise.
Consequently, to derive analytical estimates, they resort to the \emph{active} injection of probe states (e.g., GHZ states) and rely on restrictive assumptions, such as isotropic errors ($p_x \approx p_y \approx p_z$) or star topologies. 
{\em In contrast,} SCOPE avoids these simplifications.
By leveraging deep learning on passive syndrome data, we capture the full anisotropic Pauli structure and decoherence effects in general topologies without halting traffic for active probes.


\section{\bf Evaluation}
\label{sec:evaluation}

\setlength{\dblfloatsep}{3pt}

\newcommand{\w}{0.7} 
\begin{figure*}[t]
\vspace*{-0.2in}
    \centering
    \includegraphics[width=\textwidth]{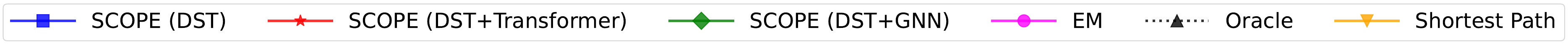} 
    \includegraphics[width=\w\textwidth]{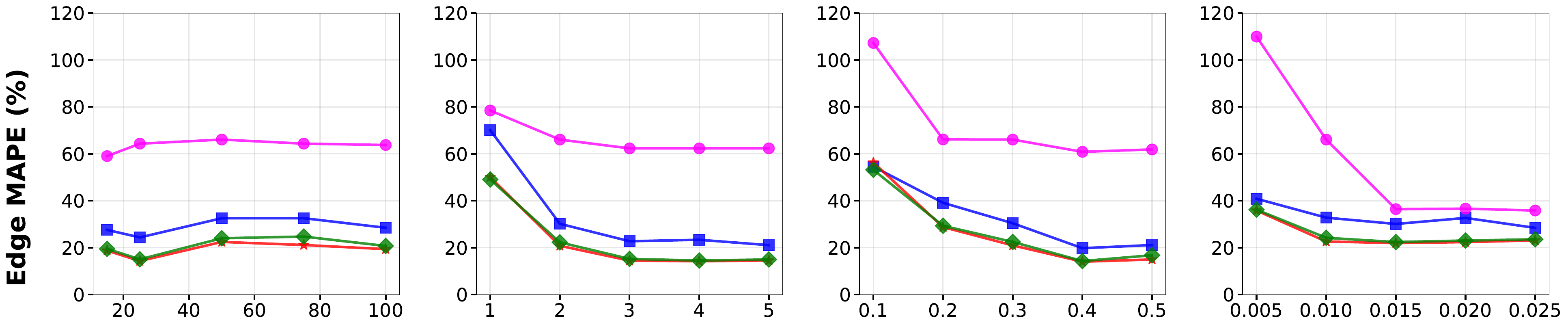}
    \includegraphics[width=\w\textwidth]{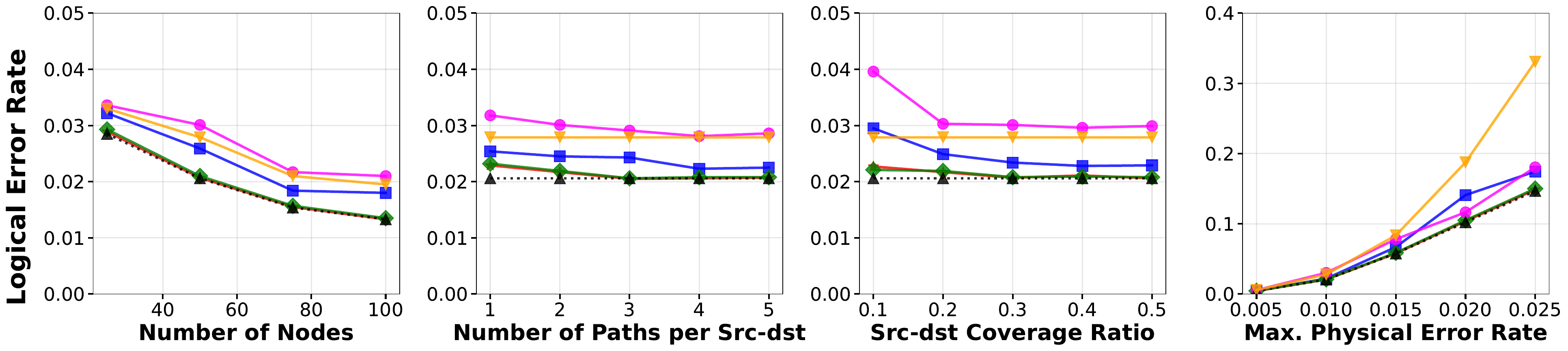}
    \vspace*{-0.1in} 
    \caption{\small Estimation error (MAPE) and logical error rate (LER) for the dependent-error and direct transmission.}
    \label{fig:dep-path}
\end{figure*}

\begin{figure*}[t]
    \centering
    \includegraphics[width=\w\textwidth]{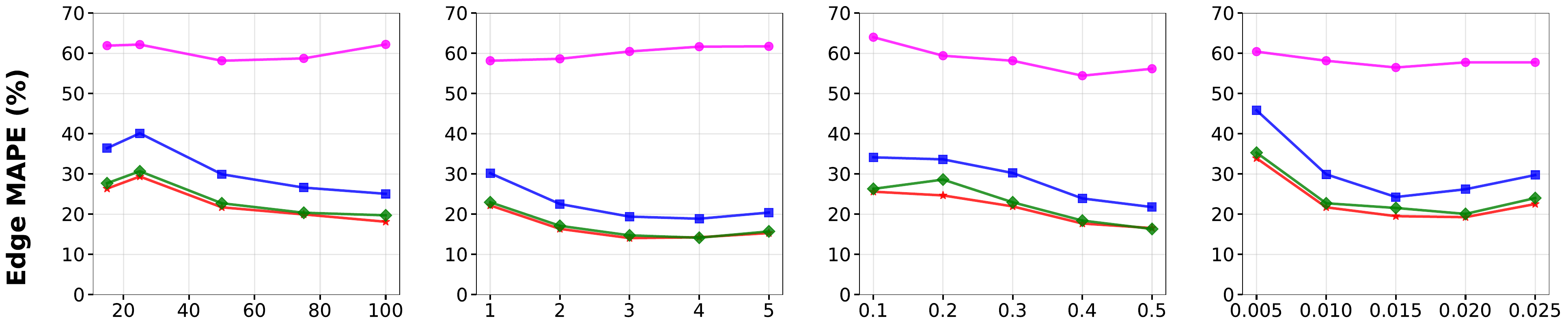}
    \includegraphics[width=\w\textwidth]{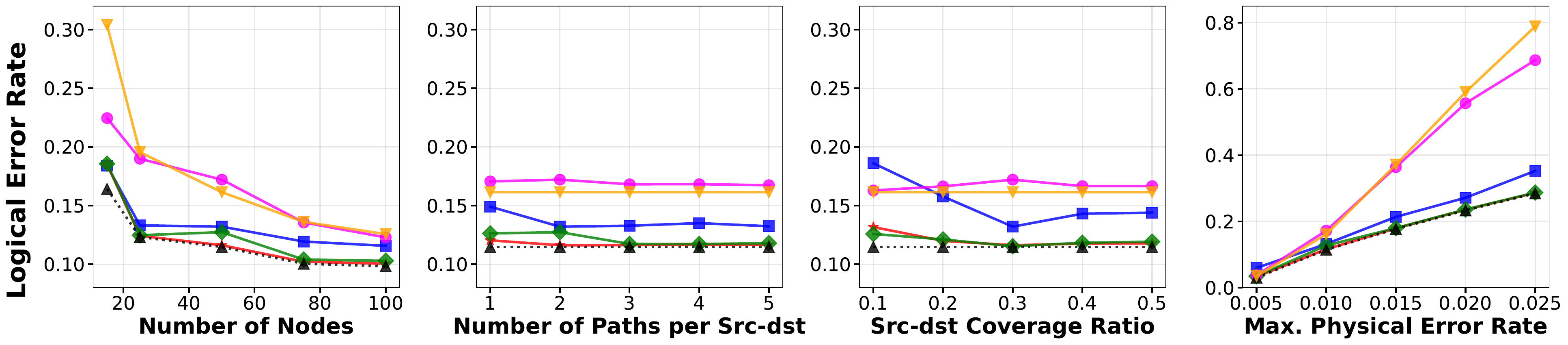}
    \vspace*{-0.1in}
    \caption{\small Estimation error (MAPE) and logical error rate (LER) for the dependent-error and teleportation-based transport.}
     \vspace*{-0.25in}
    \label{fig:dep-tele}
\end{figure*}

We evaluate \scope using the quantum network simulator NetSquid~\cite{netsquid2020} and IBM hardware-calibrated error profiles.

\para{NetSquid Protocol Implementation.}
As outlined in \S\ref{sec:gen}, we implement \scope by modeling a quantum network processing continuous background traffic.
Background requests are generated periodically at each node to sample a configurable fraction of source--destination pairs.
For each pair, transmissions are distributed over the top-$k$ shortest paths, with each path probed using a diverse set of QEC codes to accumulate syndrome statistics.
The centralized controller aggregates syndromes and path metadata from this background traffic.
Once sufficient data have been collected, the syndrome-driven inference engine updates the global error map.
Routing tables are then refreshed to map each destination to the path and code pair $(\mathcal{P}^*, \mathcal{C}^*)$ that minimizes the predicted logical error rate (LER).
For subsequent requests, sources utilize these updated tables, while new syndromes generated during operation are fed back to the controller, enabling continuous online re-estimation and adaptation.

\para{Simulation Setting.}
Unless stated otherwise, we simulate a network with an edge density of 0.2, using $N=15$ to $100$ nodes (default: $N=50$). \emph{Traffic \& Probing:} We vary the coverage ratio of active source-destination pairs from 0.1 to 0.5 (default 0.3).
For each pair, we explore $k \in [1, 5]$ shortest paths (default $k=2$) and collect {\bf 1000 shots} per path across a set of stabilizer codes $\{ \llbracket 5,1,3\rrbracket, \llbracket7,1,3\rrbracket, \llbracket9,1,3\rrbracket, \llbracket8,2,3\rrbracket \}$. 
\underline{Noise Models:} Physical noise is modeled by assigning each link and node gate a base error probability drawn uniformly from $[0, p_{\max}]$, with $p_{\max}$ varying from 0.005 to 0.025 (default 0.01). Intra-block qubit errors are applied independently.
For teleportation, we set the probabilities of successful swapping and fusion to 0.4.
Decoherence follows a standard $T_1/T_2$ relaxation model with $T_1=T_2=1$~s.
Connection requests are generated concurrently with an interarrival time of 50 ms.
\underline{Dependent Error Injection:} To simulate non-Markovian network conditions, we introduce a ``stressed state'' for network elements.
Links (in direct transmission) and node gates (in teleportation) are assigned a high-noise profile ($3\times$ baseline).
During operation, the effective error rate interpolates between the baseline and stressed states via a history-dependent variable $h \in [0,1]$, which serves as a proxy for network load or crosstalk accumulation.
$h$ increases sigmoidally when prior transmissions along the same routing path (or swapping subtree) result in errors.

\para{Metrics; Baselines.}
We evaluate performance using three metrics:
(i) \emph{MAPE} quantifies the accuracy of link-error parameter estimation. It computes the average relative deviation $|\hat{x}-x|/x$ over all links, where $\hat{x}$ is the model's estimate (averaged over contexts for dependent noise) and $x$ is the ground truth.
(ii) \emph{TVD} measures the model's predictive fidelity via the syndrome-histogram distance $\tfrac{1}{2}\|\mathbf{h}_{\mathrm{pred}}-\mathbf{h}_{\mathrm{obs}}\|_1$. Note that while accurate parameter estimation (low MAPE) implies accurate prediction (low TVD), the converse does not necessarily hold, as multiple error profiles can generate identical syndrome distributions (parameter degeneracy).
(iii) \emph{LER} (Logical Error Rate) serves as the primary performance benchmark. It represents the ground-truth probability of logical failure for the specific (code, path) configuration selected by the protocol.
We compare \scope against two \underline{baselines}. (i) \me: an expectation--maximization estimator that assumes \emph{independent} link errors and fits a single static per-link error table from end-to-end syndromes. (ii) Shortest Path (\shp, used only for LER comparisons): a topology-only routing baseline that ignores error heterogeneity and only chooses the shortest path. Finally, we include an {\tt Oracle} baseline with perfect knowledge of the ground-truth error profile, representing the optimal achievable performance.

\para{NetSquid Results.}
Figs.~\ref{fig:dep-path} and~\ref{fig:dep-tele} present MAPE and LER metrics for direct transmission and teleportation, focusing on the challenging dependent-error regime. 
In the direct transmission setting (Fig.~\ref{fig:dep-path}), standard baselines (\me and \shp) generally perform significantly worse than \scope.
 Notably, the independence-assuming \dst variant lags behind the full correlation-aware models, demonstrating the limitations of local estimation.
Quantitatively, \me suffers a high MAPE of $\approx$60\% (yielding an LER of $\approx$3\%), whereas \scope's Transformer and GNN variants achieve a MAPE of $\approx$20\% (reducing LER to $\approx$2\%).
In high-error regimes, however, \me tracks \scope closely (within 10--15\%), likely due to the diminished relative impact of inter-link correlations.
In the teleportation setting (Fig.~\ref{fig:dep-tele}), \scope maintains a robust advantage; unlike the direct case, the performance gap here remains significant across all error rates.
Overall, the Transformer and GNN architectures consistently match the optimum {\tt Oracle}.
\blue{We observe that the narrow gap between \scope and \texttt{Oracle} suggests that passive syndrome estimation recovers near-complete
error visibility without active probing.}

\para{Ablation Studies.}
\blue{We next make explicit the main ablation axes in our evaluation: the error-estimation engine and the route/code decision engine.}

\softpara{Error-Estimator Ablations.}
\blue{Figs.~\ref{fig:dep-path} and~\ref{fig:dep-tele} already isolate two estimation gains:
(i) the value of correlation-aware refinement, by comparing \scope (DST+Transformer/GNN) against \scope (DST), and
(ii) the value of our  QEC-aware differentiable estimator, by comparing \scope variants against the baseline \me.
In the dependent-error regime, DST incurs roughly 50\% higher MAPE than the full DST+Transformer/GNN model, confirming that static per-edge estimation cannot capture path-dependent noise.
Similarly, \me suffers $\approx 60\%$ MAPE, whereas \scope's correlation-aware variants achieve $\approx 20\%$.}

\begin{figure}
    \centering
    \includegraphics[width=\linewidth]{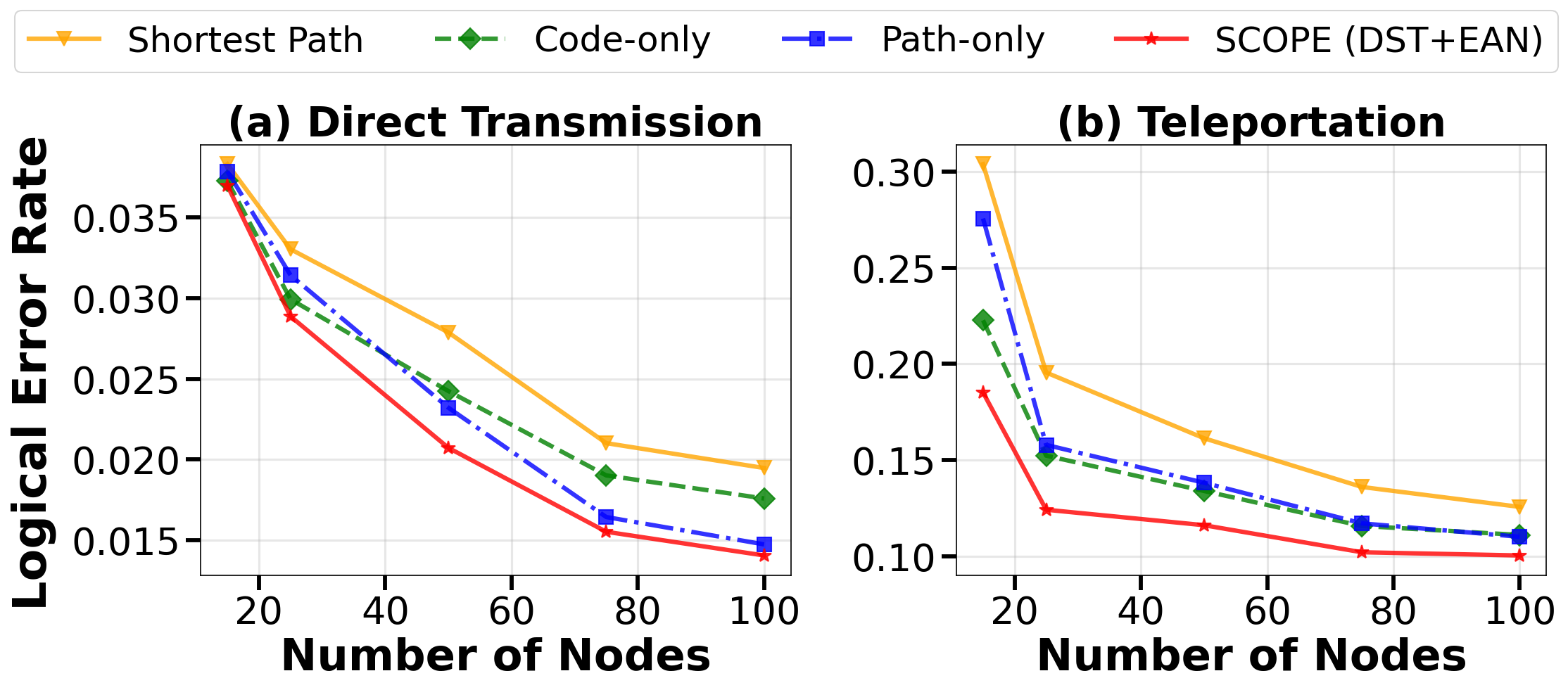}
    \caption{\small \blue{Route-code optimization ablation: LER vs.\ network size under (a) direct transmission and (b) teleportation. \eat{\texttt{Shortest Path} uses the shortest route with a fixed default code; \emph{Code-Only} fixes the route to the shortest path but selects the best code; \emph{Path-Only} fixes the code but optimizes the route; \scope jointly optimizes both.}}}
    \label{fig:ablation-routing}
\end{figure}

\softpara{Route/code optimization ablations.}
\blue{To disentangle the contributions of route selection and code selection, we evaluate two partial-optimization baselines: \emph{Code-Only} fixes the route to the shortest path but selects the best code, while \emph{Path-Only} fixes the code but optimizes the route. Full \scope jointly optimizes both path and code.
Figs.~\ref{fig:ablation-routing}(a) and~\ref{fig:ablation-routing}(b) compare the LER of these baselines against full \scope and \texttt{Shortest Path}, for direct transmission and teleportation, respectively. In both settings, each partial baseline improves over \texttt{Shortest Path}, but neither matches the performance of full joint optimization. This confirms that route and code selection provide complementary gains, and that \scope's improvements are not attributable to either decision alone.}

\begin{figure}
\vspace*{-0.2in}
    \centering
    \includegraphics[width=\linewidth]{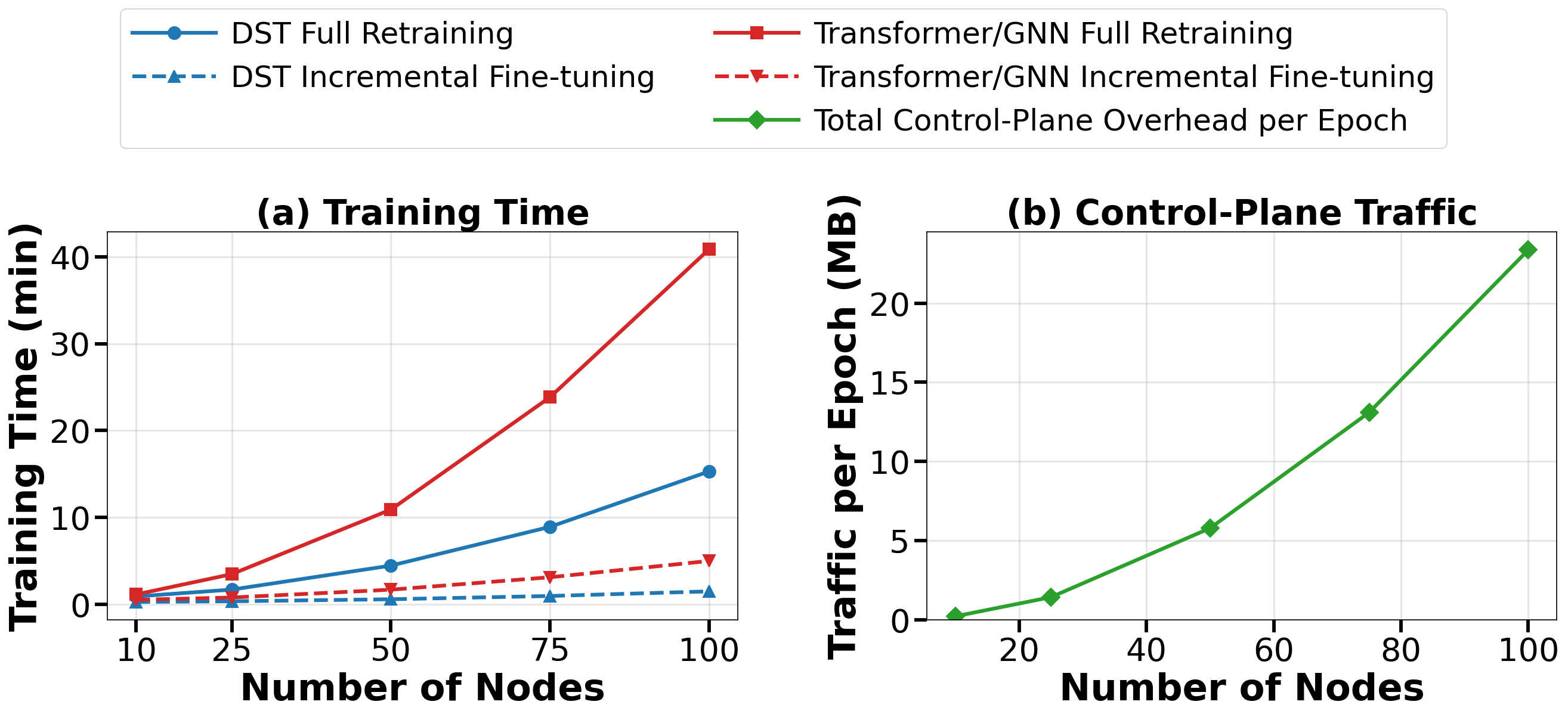}
    \caption{\small \blue{Operational overhead for varying network size.
    (a) Computational overhead: Wall-clock training time for DST
    and Transformer/GNN, for full retraining as well as incremental fine-tuning, on an RTX 4070 Ti.
    (b) Communication overhead: Total control-plane traffic per
    update epoch.}}
    \label{fig:overhead-all}
\end{figure}

\para{\blue{Overhead Analysis.}}
\blue{SCOPE's control-plane computations---error-map updates and route/code re-optimization---run asynchronously and are fully decoupled from the critical path of data-plane requests (\S\ref{sec:system}). The update frequency is dictated by the link-quality drift timescale; recent measurements of deployed quantum links report drift on the order of hours after physical-layer calibration~\cite{sena2025high}. 
Fig.~\ref{fig:overhead-all}(a) reports wall-clock training time on an RTX 4070 Ti. At $N=100$ nodes, full retraining of DST and EAN takes $\sim$15 and $\sim$40 minutes, respectively, both within the hourly drift window. 
In normal operation, however, \scope uses incremental fine-tuning, which requires only $\sim$1--5 minutes, leaving ample margin for continuous adaptation.
Fig.~\ref{fig:overhead-all}(b) shows that per-epoch control traffic, including syndrome histograms and route/code-table updates, remains below 25 MB even at $N=100$, negligible relative to conventional classical control-plane capacity.}


\begin{figure}[t]
    \centering
    \vspace*{-0.15in}
    \includegraphics[width=\linewidth]{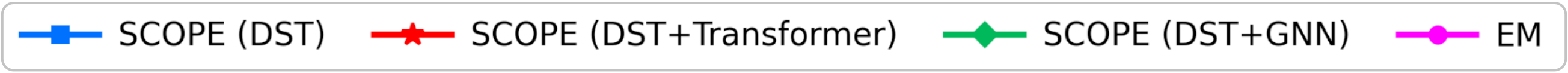}\par\vspace{-0.4ex}
    \includegraphics[width=\linewidth]{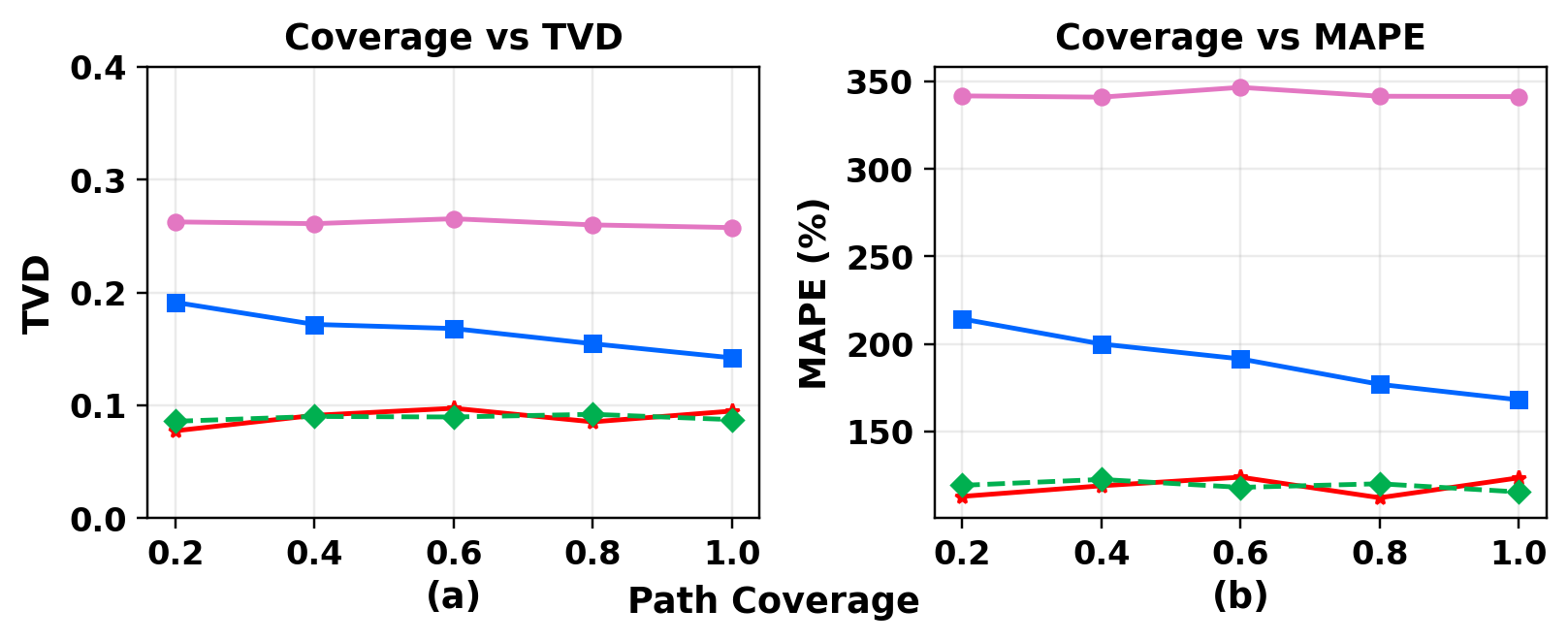} \par\vspace{-0.5ex}
    \includegraphics[width=\linewidth]{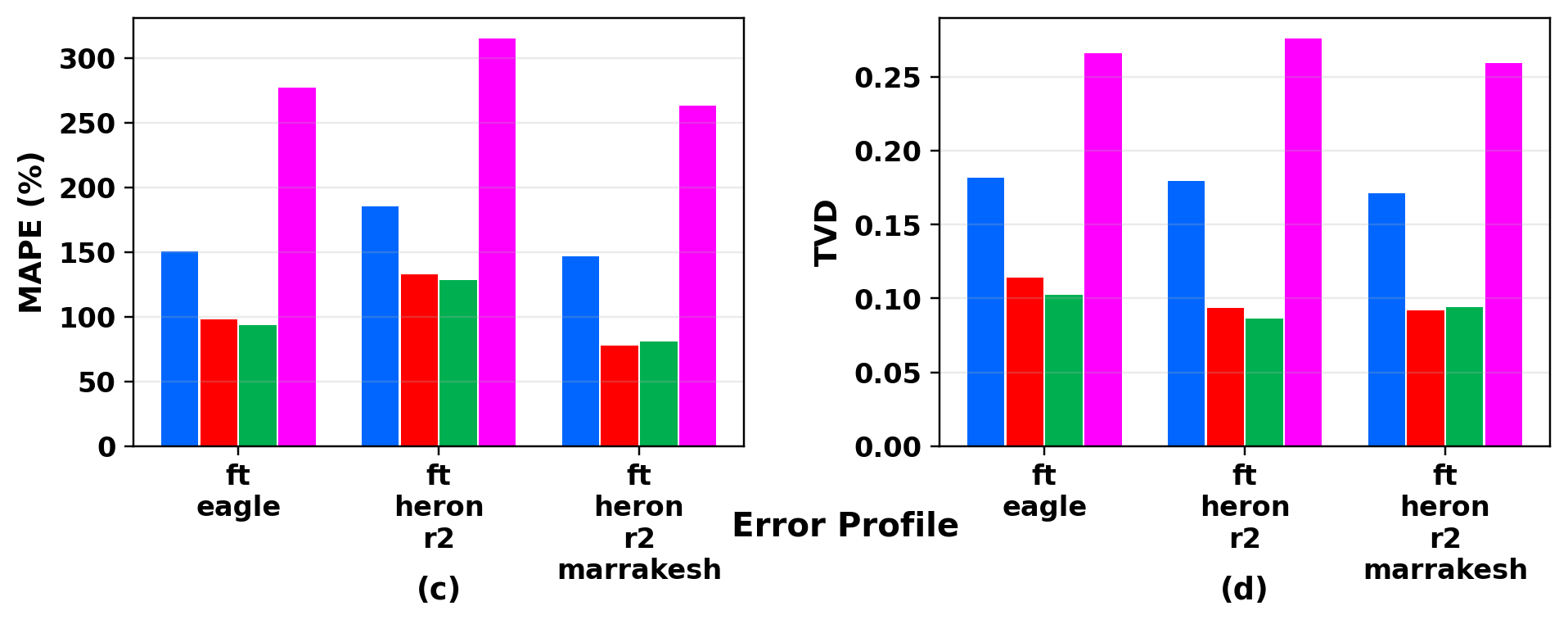}
    \vspace*{-0.2in}
    \caption{\small MAPE and TVD metrics under IBM hardware-calibrated noise. The top row shows results for varying path coverage (default profile), while the bottom row shows results for the other three hardware error architectures.}
    \vspace*{-0.25in}
    \label{fig:ibm-results}
\end{figure}
\para{Simulating on IBM Hardware-Calibrated Noise Models.} To validate \scope in the absence of physical QNs, we emulate a distributed architecture on Qiskit's \texttt{AerSimulator}~\cite{aersimulator} by partitioning qubits into logical nodes and implementing links via SWAP operations. We utilize simulation rather than physical execution to ensure: (1) \emph{Topological Independence}: The sparse coupling of current chips (e.g., heavy-hex) necessitates excessive SWAP overheads to embed logical nodes, introducing confounding errors. (2) \emph{Transport Isolation}: Physical connectivity constraints force SWAPs even for local operations, raising the noise floor. Simulation allows us to combine an ideal connectivity graph (for clean intra-node operations) with realistic, hardware-calibrated noise models (for transport links), thereby isolating the routing signal.

\softpara{Methodology.}
We evaluate four QEC codes ($\llbracket 5,1,3 \rrbracket, \llbracket 7,1,3 \rrbracket,$ $\llbracket 9,1,3 \rrbracket, \llbracket 8,2,3 \rrbracket$), allocating $n$ data and 4--8 ancilla qubits per block.
With additional link qubits for SWAP-based transport, each node requires 15--20 physical qubits. This allows us to simulate a 6-node network ($\approx 120$ total qubits) on the \texttt{AerSimulator}, with 9 links. Circuits are transpiled to native gates (\texttt{cz}, \texttt{sx}, \texttt{x}, \texttt{rz}) and executed using a hardware-derived noise model. This model encompasses gate infidelities and thermal relaxation ($T_1/T_2$), but we selectively disable readout errors to decouple measurement imperfections from transport fidelity loss. 
We evaluate performance across four distinct error profiles,  corresponding to the four processor architectures available in the IBM simulator.

\softpara{IBM Hardware Noise Results.}
We focus exclusively on direct transmission, as the simulator's lack of native support for stochastic time evolution precludes a faithful evaluation of entanglement swapping with wait times.
Fig.~\ref{fig:ibm-results} details performance metrics, varying path coverage in (a)--(b) and error profiles in (c)--(d).
While the complex, hardware-derived noise landscape results in elevated MAPE across all schemes, \scope maintains a significant advantage over the \me baseline.
Notably, TVD values remain comparatively low despite higher MAPE; this suggests that while precise link-level estimation is challenging against an independence-based ground truth, \scope successfully recovers accurate path-level error distributions.
Unfortunately, the limited path diversity of this small-scale topology saturates the routing gain; consequently, we observe that all schemes converge to the performance of the {\tt Shortest Path} baseline (LER data omitted for brevity as curves are statistically indistinguishable).

\para{\blue{Hardware Trajectory and Simulation Rationale.}}
\blue{Large-scale QEC-enabled quantum networks do not yet exist, so our evaluation uses simulation to study the regime in which SCOPE is intended to operate, while using hardware-calibrated emulation to ground the noise models in present devices. This evaluation is conservative: current hardware-scale emulations have limited path diversity and therefore understate the routing gains that SCOPE can exploit. As hardware scales, two trends favor \scope's design. First, larger networks---more nodes and links at fixed edge density---increase path diversity, giving \scope more opportunities to avoid high-error or bias-mismatched links. This trend is already visible in our NetSquid sweep (Figs.~\ref{fig:dep-path}--\ref{fig:dep-tele}), where MAPE remains stable, and LER decreases as the network grows to 100 nodes. 
Second, more qubits per node make larger or bias-tailored QEC codes feasible, increasing the value of joint route/code optimization and providing more informative syndrome telemetry through additional stabilizer checks. To keep estimation scalable, SCOPE need not learn from raw $2^r$-syndrome histograms; it can aggregate compact per-stabilizer error statistics as code size grows.}

\section{Conclusions}

In this work, we demonstrate that effective noise models learned directly from \textit{in situ} syndromes provide a superior basis for quantum network routing compared to static hardware calibrations.
Our results show that the predictive accuracy of syndrome distribution can capture the complex, context-dependent errors inherent in transport.
This shift from idealized component benchmarks to optimizing for observable network behavior is essential for maximizing near-term transmission or entanglement fidelity.
Ultimately, these models enable the co-design of network-optimized error correction codes, tailoring protection schemes to the transport fabric's specific operational fingerprints.

\bibliographystyle{IEEEtran}
\bibliography{reference}

\begin{thebibliography}{10}
\providecommand{\url}[1]{#1}
\csname url@samestyle\endcsname
\providecommand{\newblock}{\relax}
\providecommand{\bibinfo}[2]{#2}
\providecommand{\BIBentrySTDinterwordspacing}{\spaceskip=0pt\relax}
\providecommand{\BIBentryALTinterwordstretchfactor}{4}
\providecommand{\BIBentryALTinterwordspacing}{\spaceskip=\fontdimen2\font plus
\BIBentryALTinterwordstretchfactor\fontdimen3\font minus \fontdimen4\font\relax}
\providecommand{\BIBforeignlanguage}[2]{{%
\expandafter\ifx\csname l@#1\endcsname\relax
\typeout{** WARNING: IEEEtran.bst: No hyphenation pattern has been}%
\typeout{** loaded for the language `#1'. Using the pattern for}%
\typeout{** the default language instead.}%
\else
\language=\csname l@#1\endcsname
\fi
#2}}
\providecommand{\BIBdecl}{\relax}
\BIBdecl

\bibitem{muralidharan2016optimal}
S.~Muralidharan \emph{et~al.}, ``Optimal architectures for long distance quantum communication,'' \emph{Sci. Rep.}, 2016.

\bibitem{nickerson2013topological}
N.~H. Nickerson \emph{et~al.}, ``Topological quantum computing with a very noisy network and local error rates approaching one percent,'' \emph{Nat. Commun.}, 2013.

\bibitem{wehner2018quantum}
S.~Wehner \emph{et~al.}, ``Quantum internet: A vision for the road ahead,'' \emph{Science}, 2018.

\bibitem{van2013path}
R.~Van~Meter \emph{et~al.}, ``Path selection for quantum repeater networks,'' \emph{Networking Science}, 2013.

\bibitem{shi2020redundant}
Y.~Zhao and C.~Qiao, ``Redundant entanglement provisioning and selection for throughput maximization in quantum networks,'' in \emph{IEEE INFOCOM}, 2021.

\bibitem{pfister2018capacity}
C.~Pfister \emph{et~al.}, ``Capacity estimation and verification of quantum channels with arbitrarily correlated errors,'' \emph{Nat. Commun.}, 2018.

\bibitem{nielsen2000quantum}
M.~A. Nielsen and I.~L. Chuang, \emph{Quantum Computation and Quantum Information}.\hskip 1em plus 0.5em minus 0.4em\relax Cambridge University Press, 2000.

\bibitem{tuckett2018ultrahigh}
D.~K. Tuckett \emph{et~al.}, ``Ultrahigh error threshold for surface codes with biased noise,'' \emph{Phys. Rev. Lett.}, 2018.

\bibitem{ghaderibaneh2022efficient}
M.~Ghaderibaneh \emph{et~al.}, ``Efficient quantum network communication using optimized entanglement swapping trees,'' \emph{IEEE Trans. Quantum Eng.}, 2022.

\bibitem{blume2017demonstration}
R.~Blume-Kohout \emph{et~al.}, ``Demonstration of qubit operations below a rigorous fault tolerance threshold with gate set tomography,'' \emph{Nat. Commun.}, 2017.

\bibitem{chuang1997prescription}
I.~L. Chuang and M.~A. Nielsen, ``Prescription for experimental determination of the dynamics of a quantum black box,'' \emph{J. Mod. Opt.}, 1997.

\bibitem{aguado2019engineering}
A.~Aguado \emph{et~al.}, ``The engineering of software-defined quantum networks,'' \emph{IEEE Commun. Mag.}, 2019.

\bibitem{ExampleCoherent}
D.~Greenbaum and Z.~Dutton, ``Modeling coherent errors in quantum error correction,'' \emph{Quantum Sci. Technol.}, 2017.

\bibitem{SwitchPaper}
M.~Alshowkan \emph{et~al.}, ``Reconfigurable quantum local area network over deployed fiber,'' \emph{PRX Quantum}, 2021.

\bibitem{Hartmann2007Role}
L.~Hartmann \emph{et~al.}, ``Role of memory errors in quantum repeaters,'' \emph{Phys. Rev. A}, 2007.

\bibitem{Goodenough2025SwapASAP}
K.~Goodenough \emph{et~al.}, ``On noise in swap {ASAP} repeater chains,'' \emph{Quantum}, 2025.

\bibitem{vaswani2017attention}
A.~Vaswani \emph{et~al.}, ``Attention is all you need,'' in \emph{NeurIPS}, 2017.

\bibitem{scarselli2008graph}
F.~Scarselli \emph{et~al.}, ``The graph neural network model,'' \emph{IEEE Trans. Neural Netw.}, 2008.

\bibitem{dahlberg2019link}
A.~Dahlberg \emph{et~al.}, ``A link layer protocol for quantum networks,'' in \emph{ACM SIGCOMM}, 2019.

\bibitem{alshowkan2021advanced}
M.~Alshowkan \emph{et~al.}, ``Advanced architectures for high-performance quantum networking,'' \emph{J. Opt. Commun. Netw.}, 2022.

\bibitem{munro2012quantum}
W.~J. Munro \emph{et~al.}, ``Quantum communication without the necessity of quantum memories,'' \emph{Nat. Photon.}, 2012.

\bibitem{pant2019routing}
M.~Pant \emph{et~al.}, ``Routing entanglement in the quantum internet,'' \emph{npj Quantum Inf.}, 2019.

\bibitem{li2021effective}
C.~Li \emph{et~al.}, ``Effective routing design for remote entanglement generation on quantum networks,'' \emph{npj Quantum Inf.}, 2021.

\bibitem{caleffi2017optimal}
M.~Caleffi, ``Optimal routing for quantum networks,'' \emph{IEEE Access}, 2017.

\bibitem{fowler2014scalable}
A.~G. Fowler \emph{et~al.}, ``Scalable extraction of error models from the output of error detection circuits,'' 2014, arXiv:1405.1454.

\bibitem{harper2020efficient}
R.~Harper \emph{et~al.}, ``Efficient learning of quantum noise,'' \emph{Nat. Phys.}, 2020.

\bibitem{flammia2020efficient}
S.~T. Flammia and J.~J. Wallman, ``Efficient estimation of {P}auli channels,'' \emph{ACM Trans. Quantum Comput.}, 2020.

\bibitem{wagner2021optimal}
T.~Wagner \emph{et~al.}, ``Optimal noise estimation from syndrome statistics of quantum codes,'' \emph{Phys. Rev. Res.}, 2021.

\bibitem{wagner2022pauli}
------, ``Pauli channels can be estimated from syndrome measurements in quantum error correction,'' \emph{Quantum}, 2022.

\bibitem{wagner2023learning}
------, ``Learning logical {P}auli noise in quantum error correction,'' \emph{Phys. Rev. Lett.}, 2023.

\bibitem{flammia2012quantum}
S.~T. Flammia \emph{et~al.}, ``Quantum tomography via compressed sensing: error bounds, sample complexity and efficient estimators,'' \emph{New J. Phys.}, 2012.

\bibitem{sarovar2020detecting}
M.~Sarovar \emph{et~al.}, ``Detecting crosstalk errors in quantum information processors,'' \emph{Quantum}, 2020.

\bibitem{de2024quantum}
M.~G. De~Andrade \emph{et~al.}, ``Quantum network tomography,'' \emph{IEEE Network}, 2024.

\bibitem{netsquid2020}
T.~Coopmans \emph{et~al.}, ``{NetSquid}, a discrete-event simulation platform for quantum networks,'' \emph{Commun. Phys.}, 2021.

\bibitem{sena2025high}
M.~Sena \emph{et~al.}, ``High-fidelity quantum entanglement distribution in metropolitan fiber networks with co-propagating classical traffic,'' \emph{J. Opt. Commun. Netw.}, 2025.

\bibitem{aersimulator}
A.~Javadi-Abhari \emph{et~al.}, ``Quantum computing with {Q}iskit,'' 2024, arXiv:2405.08810.

\end{thebibliography}

\end{document}